\documentclass[usenatbib]{mnras}

\usepackage{mathptmx}
\usepackage{txfonts}
\usepackage{amsmath}
\usepackage{fleqn}
\usepackage{natbib}

\usepackage[T1]{fontenc}
\usepackage{ae,aecompl}
\usepackage{epsfig}

\newcommand{\co}{\mbox{$^{12}$CO}}
\newcommand{\coa}{\mbox{$^{13}$CO}}

\newcommand{\kms}{\mbox{km s$^{-1}$}}
\newcommand{\htwo}{\mbox{H$_2$}}

\newcommand{\vlsr}{\mbox{$V_{{\rm LSR}}$}}

\title[Short title, max. 45 characters]{MNRAS \LaTeXe\ template -- title goes here}

\title
[Relative orientations in the Taurus cloud]{The relative orientation between the magnetic field and gradients of surface brightness within thin 
velocity slices of \co\ and \coa\ emission from 
the Taurus molecular cloud}
\author[M. Heyer, J.D. Soler, B. Burkhart]
{M.~Heyer$^1$\thanks{Email:heyer@astro.umass.edu}, 
J.~D. Soler$^2$, 
and B. Burkhart$^{3,4}$
\\
$^{1}$Department of Astronomy, University of Massachusetts, Amherst, MA 01003, USA\\
$^{2}$ Max Planck Institute for Astronomy, K\"{o}nigsthul, 17, 69117, Heidelberg, Germany \\
$^{3}$Center for Computational Astrophysics, Flatiron Institute, 162 Fifth Avenue, New York, NY 10010, USA\\
$^{4}$Department of Physics and Astronomy, Rutgers University,  136 Frelinghuysen Rd, Piscataway, NJ 08854, USA\\
}
\date{Accepted 2020 June 15. Received 2020 June 15; in original form 2020 April 02}

\pubyear{2020}

\begin{document}
\label{firstpage}
\pagerange{\pageref{firstpage}--\pageref{lastpage}}
\maketitle


\begin{abstract}
We examine the role of the interstellar magnetic field to modulate the orientation of 
turbulent flows  within 
the Taurus molecular cloud using spatial gradients of thin velocity slices of \co\ and \coa\ antenna temperatures.
Our analysis accounts for the random errors of the gradients that arise from the thermal noise of the 
spectra.
The orientations of the vectors normal 
 to 
the antenna temperature 
gradient vectors
are compared to the magnetic field orientations that are 
calculated from Planck 353~GHz polarization data.  These relative orientations are parameterized with 
the projected Rayleigh statistic and mean resultant vector.  
For \co, 28\% and 39\% of the cloud area exhibit strongly parallel or strongly perpendicular relative orientations
respectively.  For the lower opacity \coa\ emission, strongly parallel and strongly perpendicular 
orientations are found in 7\% and 43\% of the 
cloud area respectively.  
For both isotopologues, strongly parallel or perpendicular alignments are restricted to 
localized regions with low levels of 
turbulence.  
If the relative orientations serve as an observational proxy to the Alfv\'enic Mach number then our results imply local variations 
of the Alfv\'enic Mach number throughout the cloud.
\end{abstract}

\section{Introduction}                                               
Clouds of fully molecular gas within the Milky Way and nearby galaxies are characterized by internal non-thermal 
motions with 
speeds larger than the thermal sound speed of \htwo.  These supersonic 
motions are generally attributed to chaotic turbulent 
flows within the clouds and their surrounding environments that are driven by varying mechanisms at both small 
and large spatial scales  \citep{Larson:1981}.  
If the interstellar magnetic field that threads through the 
clouds is sufficiently strong and the 
collisional time scale between ions and neutral molecules is short with respect to an Alfv\'en wave crossing 
time, then the character of the turbulent flows is necessarily modified \citep{Mouschovias:2011}.  
In these conditions, both energy and momentum can be redistributed by Alfv\'en waves whose amplitudes are largest when the wave vectors
are parallel (transverse) or perpendicular (magnetosonic) to the local magnetic field 
direction.  The action of such waves introduces
anisotropies in the distribution and motions of the gas
\citep{Arons:1975, Mouschovias:2011}. 

Detailed predictions of anisotropy have emerged from theories of interstellar MHD turbulence.  
\citet{Goldreich:1995} (hereafter, GS95) 
 examined the effects of the interaction between 
oppositely-directed packets of shear Alfv\'en waves on a magnetized fluid. In the limit of strong Alfv\'enic turbulence,
they showed that turbulent eddies become elongated along 
the magnetic field direction and that the wave vectors parallel 
and perpendicular to the local magnetic field direction scale as k$_\parallel \propto k_\perp^{2/3}$. 
\citet{Goldreich:1997} extended this study to an intermediate turbulence regime and considered the 
effects of pseudo Alfv\'en waves that correspond to slow magnetosonic waves in the incompressible limit.
They demonstrated that the pseudo Alfv\'en waves (hereafter slow mode) exhibit the same velocity 
spectrum and anisotropy as 
the shear Alfv\'en modes.

The anisotropy considered in GS95 was in relation to the mean magnetic field. In fact, the 
anisotropic eddies are aligned with the direction of the local magnetic field on the scale of the eddy.  
The increasing elongation of eddies with decreasing scale in the local frame of reference was affirmed 
by \citet{Lazarian:1999} (LV99) who considered the effects of magnetic reconnection on the gas motions 
and field structure. LV99  showed that, in the presence of turbulence, the altered magnetic field line 
topology is a natural consequence of the turbulent cascade while, at the same time, reconnection itself 
can further drive turbulence. As the turbulent cascade proceeds perpendicular to the magnetic field, 
reconnection becomes fast, with reconnection taking place on order of the local eddy turnover time.  
The result of fast reconnection is that the velocity fluctuations will be aligned with the local field, 
a result which has major implications for many astrophysical processes such as cosmic ray diffusion, 
star formation and ISM structure. 
Using MHD simulations of compressible gas, \citet{Cho:2003}
further verified velocity anisotropy imposed by the magnetic field in the local frame of reference 
for both Alfven and slow modes in the regime where the ratio of thermal to magnetic field 
energy densities is $<<$1.

In addition to the MHD waves, magnetically-induced velocity
anisotropies can also emerge from the large scale flows in a
compressible medium.
For a strong magnetic field, the MHD equations predict a binary state in which gas motions are either along or 
perpendicular to the magnetic field \citep{Soler:2017}.  Furthermore, collapse of material along field lines can 
lead to hydrodynamic shocks and anisotropic density distributions \citep{Chen:2015,Mocz:2018}.

Can such magnetically-induced velocity anisotropy be measured in the molecular interstellar medium?  
The study of \citet{Cho:2003} provides a valuable 
basis on which to build a framework of analyses that can identify such conditions. 
In their set of MHD simulations, they show iso-contours of the two-dimensional structure function of 
velocities with respect to the local magnetic field direction.  As the structure function quantifies the 
degree of velocity correlation along 2 orthogonal axes, its shape and orientation represents the 
spatial configuration of 
the turbulent eddies in the reference frame of the local field.

\citet{Esquivel:2005} and later, \citet{Esquivel:2011} and \citet{Burkhart:2014} examined shapes 
and elongations of the two-dimensional 
structure functions of density-weighted velocities from a set of MHD simulations spanning a large range 
in both 
Alfv\'en and 
sonic Mach numbers.  
These studies found alignment of elongated structure functions and hence, turbulent eddies,  with 
the projected, perpendicular component of the magnetic field along the line of sight.  This alignment held
for models with sub-Alfv\'enic 
motions and sonic 
Mach numbers as high as 7. 
\citet{Esquivel:2015} extended this analysis to line intensities 
of position-position-velocity 
data derived from MHD simulations for varying magnetization and spectral resolutions.  For high spectral resolution
slices of the data cube, 
 they found 
elongated structure functions with long axes aligned with the orientation of the magnetic field for 
strongly magnetized models.  

\citet{Heyer:2008} applied 
Principal Component Analysis (PCA) to position-velocity slices of spectroscopic data cubes to 
evaluate the velocity structure function along orthogonal axes. Using model CO 
line profiles derived 
from density and velocity fields of MHD simulations, they also found the orientation of velocity 
anisotropy to be aligned with the local 
magnetic field direction but 
only for sub-Alfv\'enic models.  Applying this method to a localized 
area in the Taurus molecular cloud with striations of CO emission oriented along the 
magnetic field, they found the strongest anisotropy of the velocity structure functions is
 also aligned with the 
magnetic field.  A followup study by \citet{Heyer:2012} over a mosaic of fields in the Taurus cloud 
showed that velocity anisotopies are limited to low column density regions of the cloud.  They suggested 
that a transition from  sub-Alfv\'enic conditions in the low column density envelope of the cloud to 
super-Alfv\'enic motions in the high density cloud interior to account for the different behavior.

While structure function methods and PCA offered initial clues to MHD anisotropy, 
both these analyses are limited to analyzing 
sufficiently large areas within a cloud to 
accurately evaluate the velocity structure function along 
each axis.  Therefore, these methods are insensitive 
 to small scale anisotropies that are predicted by theories of 
MHD turbulence.

More recently, the orientation of spatial gradients of column densities and velocity centroids
 with respect to that of the 
magnetic field has been an effective tool to investigate MHD turbulence. 
\citet{Soler:2013} found a bimodal distribution of relative orientations of column density gradient vectors 
and magnetic field orientations on a set of numerical simulations with varying magnetization.  
For low density regions within the simulations and higher magnetizations, the angles of the vectors  
normal to the gradient vectors are aligned with the local magnetic field.  For higher density regions, 
the normal vectors are perpendicular to the magnetic field orientations. 
This bimodal effect is also present in a set of local clouds observed by the Planck mission, which provides 
both dust column density and dust polarization \citep{Planck:2015_XXXV}. 
\citet{Gonzalez:2017} extended this analysis to gradients of velocity centroids of spectral line 
emission (VCG) that can be more directly compared to the kinematics of MHD turbulence.  

These applications of the gradient techique analyzed either direct two-dimensional images of 
clouds such as column density 
or projections 
of spectroscopic data cubes of line emission such as moment maps.  
However, there is much more information within 
the full data cubes of spectral line emission that can be exploited.  
\citet{Lazarian:2018b} adapted the gradient 
 method to spectral line 
intensities within thin velocity slices of the data cube (Velocity channel gradients, VChG).  
For very thin spectral channels, 
the structure of line emission is most affected by the velocity field of the 
volume rather than the distribution of gas densities \citep{Lazarian:2000, Brunt:2004}.  
The advantage 
of this analysis over previous methods is its ability to define a vector that describes the magnitude and direction 
of line intensity differences that arise from the turbulent flow of gas, gravity,
or feedback processes from newborn stars  at the resolution 
of the observations. 

\citet{Hu:2019} applied the VChG method to optically thin, molecular line emission from several nearby molecular 
clouds including the Taurus \coa\ data analyzed in this study. 
They found good alignment of the magnetic field orientation with that of the vector normal to the velocity centroid gradient vector.
From the width of the distributions of either gradient angles or polarization orientations ($\delta\theta$) 
and 
applying the Davis-Chandrasekhar-Fermi relation M$_{\rm A}$ = tan($\delta\theta$) 
\citep{Davis:1951, CF:1953}, 
they estimated Alfv\'en Mach numbers in these clouds 
between 0.7 to 1.2. 

In this contribution, we examine the orientations of gradients of \co\ and \coa\ intensities within a large 
set of narrow 
velocity channels with respect to magnetic field orientations as measured 
by Planck \citep{Planck:2015_XIX} for 
the Taurus molecular cloud.  
Analysis of both optically thick \co\ emission and mostly 
optically thin \coa\ emission may provide insight to the effects of line opacity on the 
on the VChG method.
Our calculations of intensity gradients specifically 
include the effects of thermal noise 
of the spectral line data and the impact of the resultant gradient angle uncertainties on circular statistics 
used to parameterize the degree of orientation alignment.  
In \S2, we summarize our data.  
Our assumptions and analysis methods are  described in \S3.  The results of the analysis are shown 
in \S4.  In \S5, we examine the dependence of the degree of relative orientations with 
the local gas properties. The interpretation of these results are described in \S6.

\section{Data}

\subsection{$^{12}$CO and $^{13}$CO J=1-0 emission}
All molecular line data used in this study are part of the FCRAO Survey of the Taurus molecular cloud described 
by \citet{Narayanan:2008} and \citet{Goldsmith:2008}. The survey imaged \co\ and \coa\ J=1-0 emission from 96~deg$^2$
of the Taurus cloud with the 14~meter telescope of the Five College Radio Astronomy Observatory using On-the-Fly (OTF) 
mapping. 
The front-end consisted of the 32-element focal plane array receiver, SEQUOIA, that fed a set of 
autocorrelation back end spectrometers configured with a spectral resolution of 25 kHz or 0.063 \kms\ and 0.067 \kms\ 
for \co\ and \coa\ respectively.  The half power beam widths for \co\ and \coa\ are 45\arcsec\ and 47\arcsec.  The spacing 
between pixels in the final regridded map is 20\arcsec\ for each transition. 

\subsection{Polarized thermal dust emission}
We analyze the same Planck HIFI polarization data at 353~GHz that was used by \citet{Planck:2015_XXXV} to map 
the magnetic field configuration in the Taurus cloud. We assume that the orientation of the plane-of-the-sky magnetic field 
is perpendicular to the orientation of the linear polarization at 353~GHz, that is, assuming perfect dust grain alignment, 
as justified by the results of \citet{Planck:2018}.
This data set, comprised of Stokes I, Q and U maps, is smoothed to 10\arcmin\ to achieve
polarization signal-to-noise ratios $P/\sigma(P) >$ 3.
These maps are obtained using a gnomonic projection of the publicly-available whole-sky maps, 
available through the Planck Legacy Archive \footnote{\url{https://pla.esac.esa.int/}} in {\tt Healpix} format.
The initial projection is made in Galactic coordinates and subsequently transformed into equatorial coordinates 
and aligned with the $^{12}$CO and $^{13}$CO maps using the reproject package in astropy and the 
corresponding rotation of Stokes $Q$ and $U$ to this reference frame using the {\tt polgal2equ} 
included in the {\tt magnetar} package \footnote{\url{https://github.com/solerjuan/magnetar}}.

The polarization orientation of the E-vector is calculated for each position in the reprojected $U$ and $Q$ images 
following the IAU convention in which positive angles are measured counterclockwise from the positive declination axis, such that,
\begin{equation}
\Psi_{E}(\alpha,\delta) = 0.5 \arctan{\biggl(\frac{-U(\alpha,\delta)}{Q(\alpha,\delta)}\biggr)}
\end{equation}
where $\alpha,\delta$ are the equatorial coordinates and $\Psi_{E}$ is the polarization angle in 
equatorial coordinates.
Assuming elongated dust grains that are responsible for the polarization are aligned with the magnetic field,
$\Psi_{B}(\alpha,\delta) = \Psi_{E}-\pi/2$ is the orientation of the magnetic field component 
perpendicular to the plane of the sky.

\subsection{Column Density derived from thermal dust emission}
To evaluate column density, we use the same map of the 353~GHz opacity that was constructed from all-sky Planck 
intensity measurements at 353, 545, and 857~GHz and supplemented by IRAS observations at 100\;\micron\ and analyzed by 
\citet{planck2013-p06b}.  The opacity was 
smoothed to 10\arcmin\ resolution and converted into column density, $\tau_{353}/N_H = 1.2{\times}10^{-26}$ cm$^{2}$, 
assuming a dust opacity based on extinction measurments towards quasars.  The column density image was transformed 
into equatorial coordinates and similarly aligned with the CO data with the {\tt astropy reproject} package. 

\subsection{Wide field views of molecular gas and magnetic field orientations in Taurus}
\begin{figure*}
\begin{tabular}{c}
\\
\\
\epsfig{file=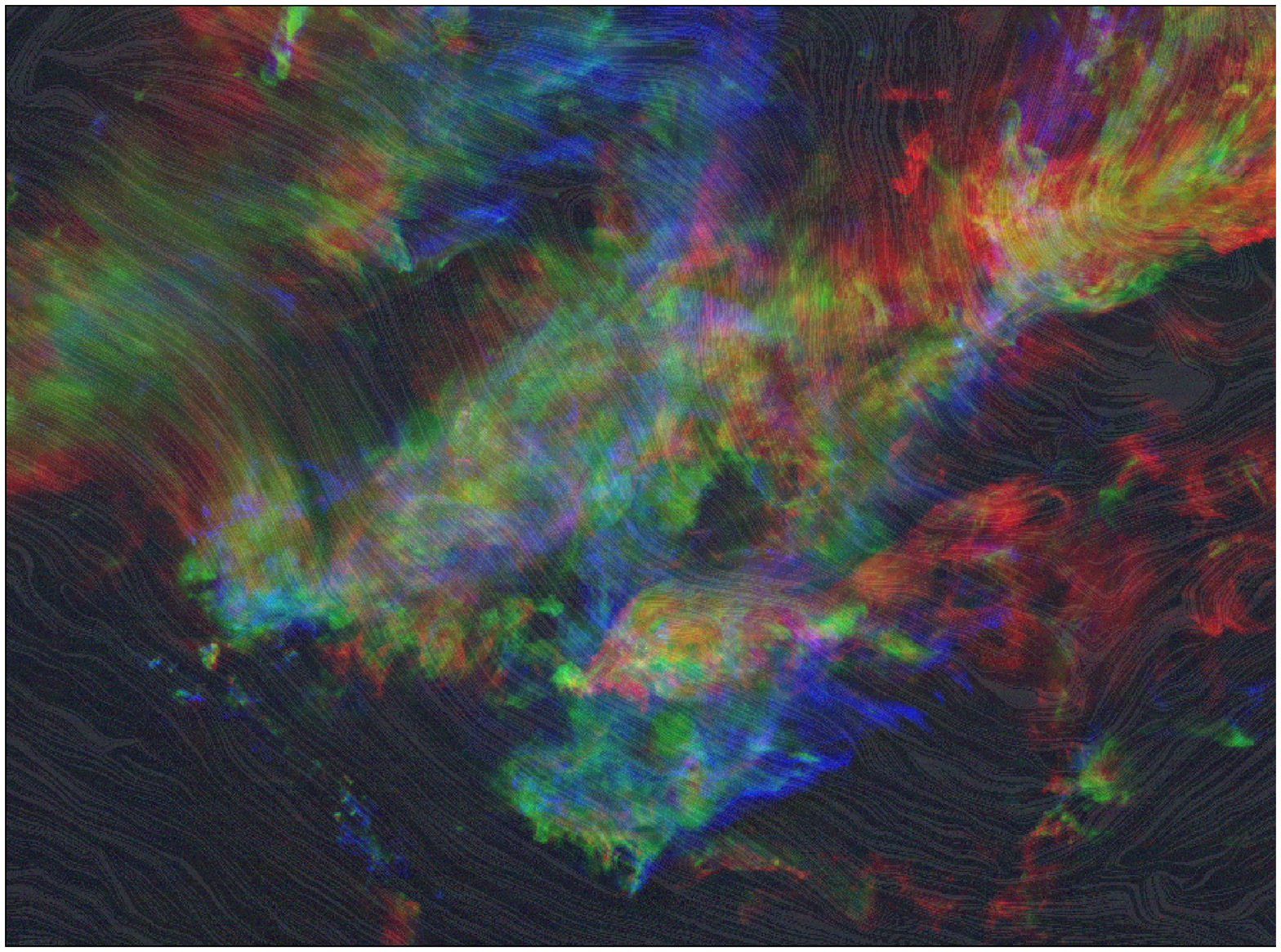,width=14cm,clip=}\\[0.5in]
\epsfig{file=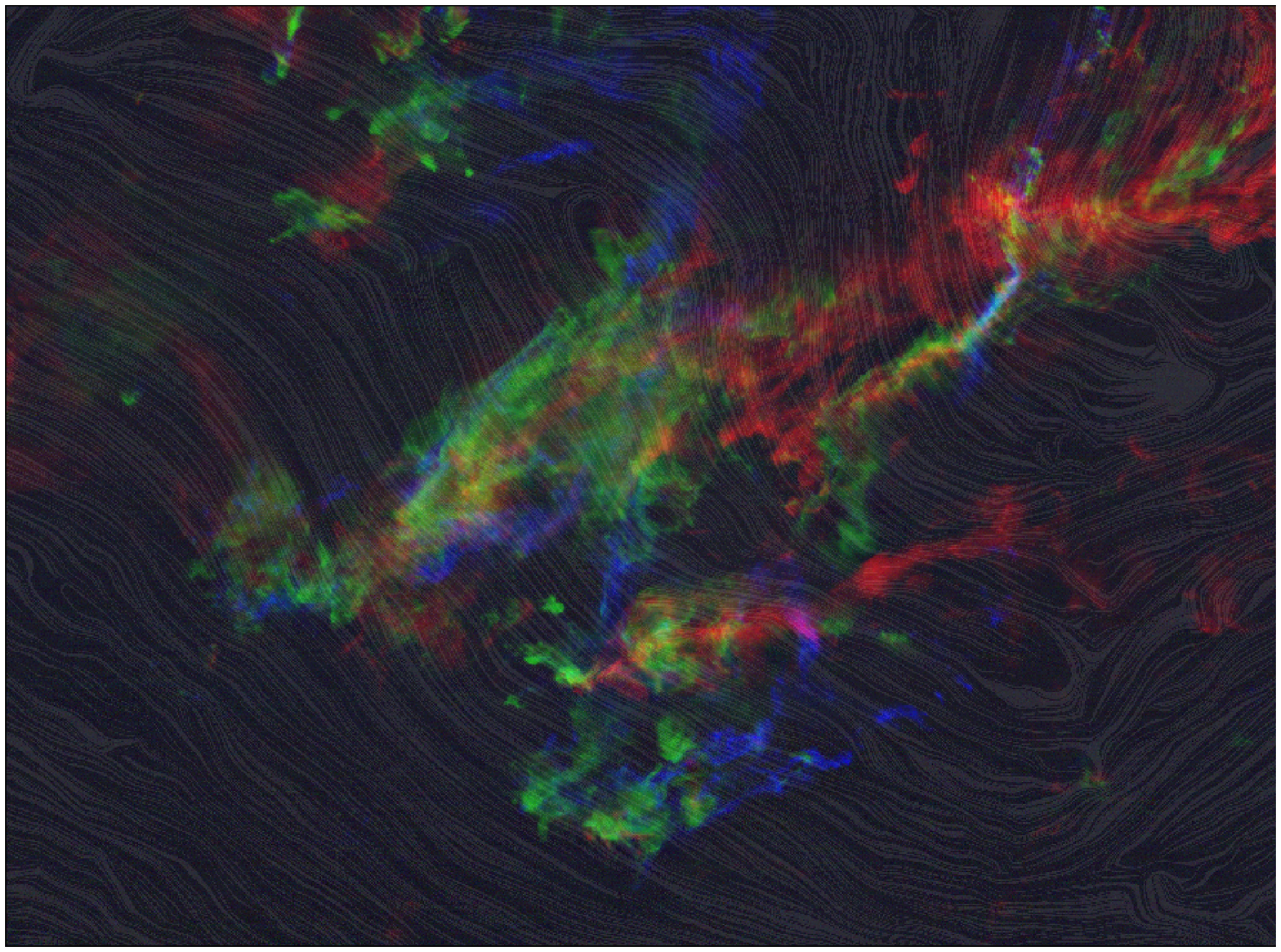,width=14cm,clip=}
\end{tabular}
\caption{Velocity-encoded RGB images of \co\ (top) and \coa\ (bottom)
 J=1-0 emission with a drapery pattern showing the magnetic field orientations 
from Planck 353~GHz polarization data.  The colors represent CO emission integrated over 
\vlsr\ intervals: 3.7-5.5 \kms (blue), 5.5-6.3 \kms (green), and 6.3-7.8 \kms (red). 
}
\label{figure1}
\end{figure*}
The wide field distribution of \co\ and \coa\ J=1-0 emission within 3 velocity intervals and a 
representation of the magnetic field orientations derived from the Planck 
353~GHz polarization data are displayed in Figure~\ref{figure1}. 
The rgb colors represent the emission integrated over velocity intervals 
through which 
most of the cloud CO luminosity is radiated: 3.7-5.5 \kms (blue), 5.5-6.3 \kms (green), and 6.3-7.8 \kms (red).  
The magnetic field orientation is imprinted 
as localized folds of the CO intensity image as produced by the line integral 
convolution method
\citep{cabral:1993}.  

The two CO images show very different intensity distributions. 
The \coa\ image is similar to the Planck 
column density map and images of visual extinction \citep{Pineda:2010} in which most of the emission is 
distributed into filaments or localized clumps.  In contrast, the \co\
image is more extended with flocculent features.  These differences arise
from the relative opacities of the two emission lines.  The high opacity of \co\ for a given velocity 
interval limits the depth to which emission can probe.  The high optical depth also maintains 
the excitation temperature by radiative trapping 
in regions where the volume density is much less than the critical density 
and allowing the \co\ line to be detected in this regime, which accounts for its more extended surface 
brightness distribution.

The magnetic field orientation varies smoothly across the cloud area.  There are no abrupt changes in the 
field orientation when crossing from a low column density area into a high density filament.  Visually,
the magnetic field aligns with the faint, wispy striation features in the northeast sector of the CO maps. 
Towards the central region of the cloud,
higher column density material is distributed perpendicular to the field orientation.  
Our goal in this investigation is to quantify these orientations in order to evaluate the more 
general alignment of the interstellar magnetic field and the gas distribution.

\section{Assumptions and Analysis}
\subsection{Thin-Slice intensity images}
Spectroscopic imaging of low J \co\ and \coa\ line emission exhibits very complex structure that results from both density variations and 
a turbulent velocity field.  \citet{Lazarian:2000} examined the impact of density and velocity on the power spectrum of 2-dimensional 
channel images of spectroscopic data cubes, $T(x,y,v)$ with varying velocity slice thickness.  In the regime for which the velocity width
of the slice, ${\Delta}v$ is greater than the velocity dispersion, $\sigma_v(L)$ over size $L$, 
the power spectrum of intensity 
fluctuations of an optically 
thin line reflect
column density variations.  
Conversely, for ${\Delta}v << \sigma_v(L)$, corresponding to the thin-slice limit, the power spectrum of 
an image of brightness temperatures is dominated by velocity variations.  Velocity Channel Analysis (VCA) developed by \citet{Lazarian:2000} is a procedure 
to separate the dependence of the power spectrum on column density and velocity 
fluctuations using both thick and thin slice 
channel widths in order to derive the functional form of the power spectrum for both 
column density and velocity fields.
In this study, we assume that structure in the thin slice channels reflect caustics 
of the
turbulent velocity field rather than density fluctuations of a turbulent cell.
This is a reasonable assumption for cold, molecular gas where the linewidth is dominated by turbulent 
broadening rather than thermal broadening \citep{Clark:2019}.

\subsection{Gradients of surface brightness in thin-slice velocity channels}
Spatial gradients of spectral line intensities, 
column densities, or 
centroid velocities are powerful measures of cloud structure and anisotropies.  
The gradient calculation produces a vector at each sampled position, (x,y), 
that describes the magnitude and direction of localized differences of field values.  In this study, we investigate the spatial gradients of 
brightness temperature of \co\ and \coa\ J=1-0 line emission in 
thin slice channels.  For notation, we define the $k$-th velocity slice of the data cube of 
brightness temperatures $T(x,y,v_k)$ as $T_k(x,y)$.  The gradient magnitude
and direction for this $k$-th slice are 
\begin{equation}
\nabla{\rm T_k(x,y)} =\bigg\{ \bigg (\frac{ \partial{ {\rm T_k(x,y)} } } {\partial{ {\rm x} } } \bigg )^2 + 
                            \bigg (\frac{ \partial{ {\rm T_k(x,y)} } } {\partial{ {\rm y} } } \bigg )^2 \bigg\}^{1/2}
\end{equation}

\begin{equation}
\Psi_{G,k} = \arctan\left[\frac{\partial T_{k}(x,y)} {\partial x} \middle/ 
                           \frac{\partial T_{k}(x,y)} {\partial y} \right] 
\end{equation}
where $\Psi_{G,k}$ is the direction of the gradient measured counter-clockwise from the $y$ (declination) axis. 
Here, we calculate the angle of the vector normal to the gradient vector,
$\Psi_{N,k} = \Psi_{G,k} + \pi/2$, which describes
the orientation of the longer axis of an emission feature that is responsible for the gradient.
Following \citet{Soler:2019}, the multi-dimensional filter routine, {\tt gaussian\_filter}, in the python {\tt scipy.filters}
 package is 
applied to the image.  
This filter includes a
smoothing option that both increases the signal to noise of the gradient measurement and 
sets an angular
 scale over which the gradient is derived, which corresponds to the Gaussian derivatives introduced by \citet{Soler:2013}. 
For the Taurus CO data, a Gaussian kernel with a full width half maximum of 8 pixels 
(2.67~\arcmin\ $\approx$ 1/4 Planck Stokes $U$ and 
$Q$ FWHM angular resolution) is applied to derive the gradient vectors.

Random noise can have a significant effect on the gradient calculation so it is important to track how noise 
in the data impacts the 
accuracy of the gradient magnitude and gradient angle.  We adopt a Monte Carlo approach to assess these errors. 
Derivatives of the thin-slice intensity images are calculated from 
$T_k(x,y)=T_{obs,k}(x,y)+T_{noise,k}(x,y)$, where $T_{obs,k}(x,y)$ is the observed antenna temperature at the position $(x,y)$ 
 and velocity slice k
and $T_{noise,k}(x,y)$ is randomly drawn value 
from a gaussian probability density distribution with the same width as 
the rms uncertainty of the spectrum at position x,y. 
The 
gradient amplitude and angle are derived from equations 1 and 2 respectively.  
This is repeated for 1024 realizations producing distributions
of the gradient amplitude and angle for each sampled position.  
The respective uncertainties are assigned to the standard deviation of each distribution for position (x,y).  

\subsection{Directional Statistics}
The gradient analysis generates 
a set of angles that are compared to the magnetic field orientation in the vicinity of the 
measurements.  
We define $\Phi_k(x,y)=\Psi_{N,k}(x,y)-\Psi_{B}(x,y)$, where 
$\Psi_{B}(x,y)$ is the angle of orientation of the interstellar magnetic field component 
perpendicular to the plane of the sky at position (x,y) 
that is interpolated from the Planck data to the 2.67' sampling interval. 
The set of $\Phi$ values is represented by the projected Rayleigh statistic, 
\begin{equation}
Z_{k} = \frac{\sum_i w_{k}(x_i,y_i) \cos(2\Phi_{k}(x_i,y_i)) } { (\sum w_{k}(x_i,y_i)^2)^{1/2} / 2}
\end{equation}
and its variance
\begin{equation}
\sigma(Z_{k})^2 = \frac {2 \sum_i (\cos(2\Phi_{k}(x_i,y_i))^2 - Z_{k}^2}{n_{ind}} 
\end{equation}
where the sums are over all sampled pixels in the area, 
$w_{k}(x_i,y_i)$ is a weighting factor for the {\it i-th} sampled position, and $n_{ind}$ 
is the number of independent measurements \citep{Jow:2018}.  It 
is evident that $Z_{k} >> 0$ for strongly aligned features, 
$Z_{k} << 0$ for features orthogonally oriented, and $Z_{k} \sim 0$ for a set of random, 
uniformly distributed orientations. 
Such random, uniform distributions can arise from a field with just noise (no signal) or a field with signal 
and significant gradient orientations but are randomly distributed or a mixture of both conditions.

In calculating the projected Rayleigh statistic, it is necessary to evaluate the number of 
independent samples as oversampling 
a field can introduce a strong bias to the result.  \citet{Fissel:2019} proposed that the number of 
independent data samples in a field is 
\begin{equation}
n_{ind}=\frac{n_{eval}}{\sigma(Z_{WN})^2} 
\end{equation}
where $n_{eval}$ is the number of evaluated pixels and $\sigma(Z_{WN})$ is the standard deviation of $Z_k$ values 
derived from a series of images filled with white noise (WN) that is smoothed and sampled identically to the data. 

The point spread functions of the Planck 353~GHz $U$ and $Q$ images 
have a full width half maximum width of 10\arcmin.  
The gradient analysis requires one to sample at finer scales than the Planck beam size in order to 
detect small scale differences in the velocity slice images. 
In our study, the thin-slice channel images are smoothed to 2.67\arcmin\ resolution
as part of the gradient calculation.
When deriving $\Phi_{N,k}$, we evaluate the gradient angle image at this same interval 
rather than the 20\arcsec\ pixel 
spacing of the CO data.  
We calculate $\sigma(Z_{WN})$ using 2 independent methods.  The first method follows the 
prescription of \citet{Fissel:2019}
in which an image is populated with spatially uncorrelated, normally distributed random values with the same dispersion 
as the data.  The image  of $\Phi_{WN}$ values derived from white noise fluctuations 
is sampled identically to the Taurus data to derive the projected Rayleigh statistic and its variance. 
This is repeated for 1024 realizations to derive a mean $Z_{k}$ value and standard deviation, $\sigma(Z_{WN})$. 
The second method evaluates the projected Rayleigh statistic in velocity channels with no signal.  This allows one to 
assess the impact of
correlated noise in our data due to On-the-Fly mapping. 
Both methods produce $Z_{k}$ values near zero and $\sigma(Z_{k}) \approx$ 1 as expected for a set of independent measurements.  
From these noise results, we conclude the oversampling introduced by the Planck beam, by a factor 
close to 16 in area, does not introduce a systemic bias in our derived values of $Z_k$ such that 
$n_{ind} = n_{eval}$ and no corrections are required.

\section{Results}
The Taurus molecular cloud presents a range of gas column densities and magnetic field orientations.  
To assess 
any dependency of the gas-field coupling on the local conditions, 
we partition the Taurus data into 150 subfields.  Each subfield is comprised of 256$\times$256 pixels (1.42$\times$1.42 deg$^2$,
which corresponds to 
physical dimensions 
3.5$\times$3.5~pc$^2$).  The subfields are spaced 
by 128 pixels in each direction.  This partitioning is identical to that applied by \citet{Heyer:2012}.
In \S8.3, we describe the effects of decreasing the subfield sizes to 128$\times$128 pixels and 64$\times$64 pixels. 

\begin{figure*}
\begin{tabular}{c}
\\
\\
\epsfig{file=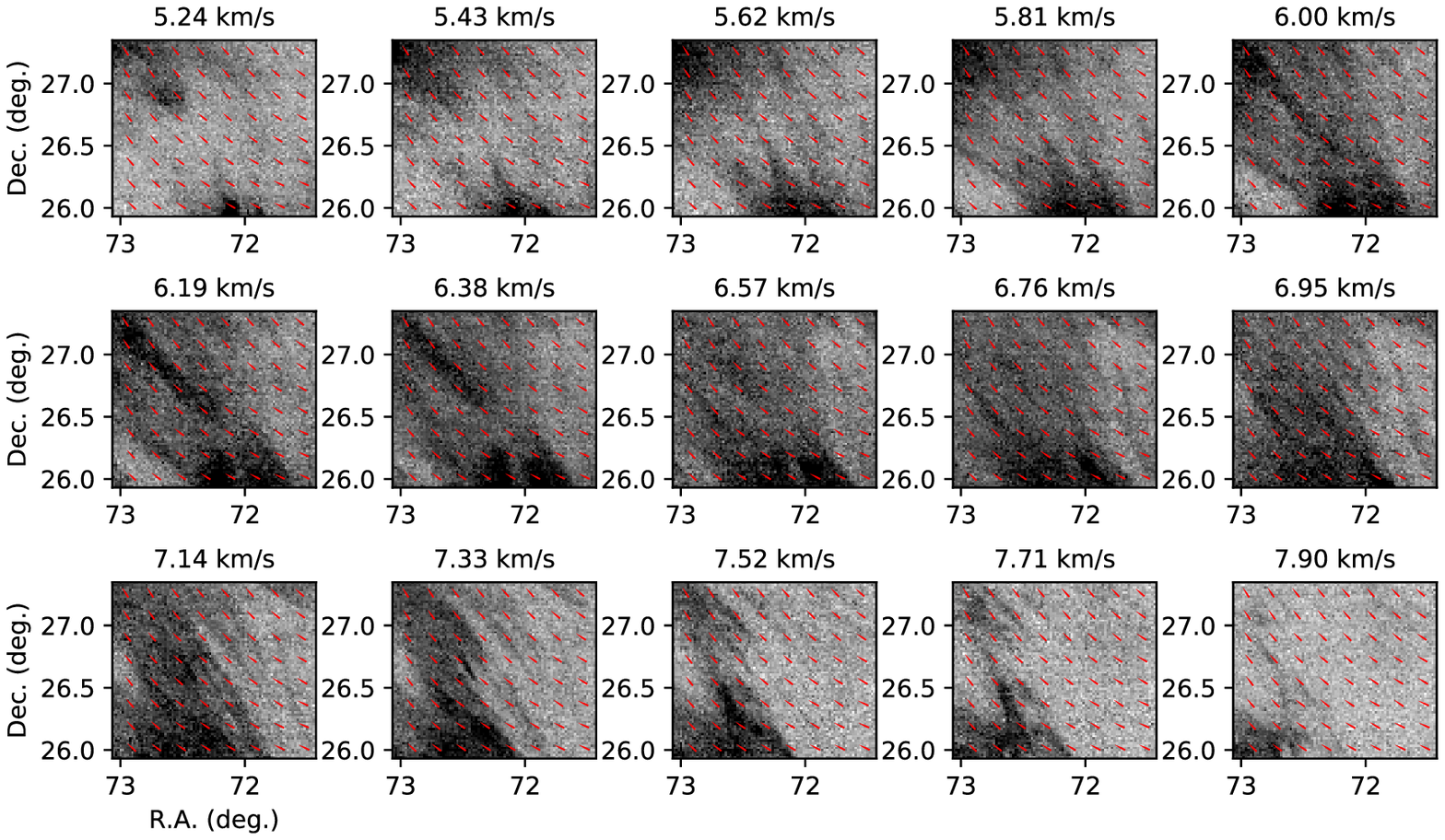,width=17cm,clip=}\\[0.5in]
\epsfig{file=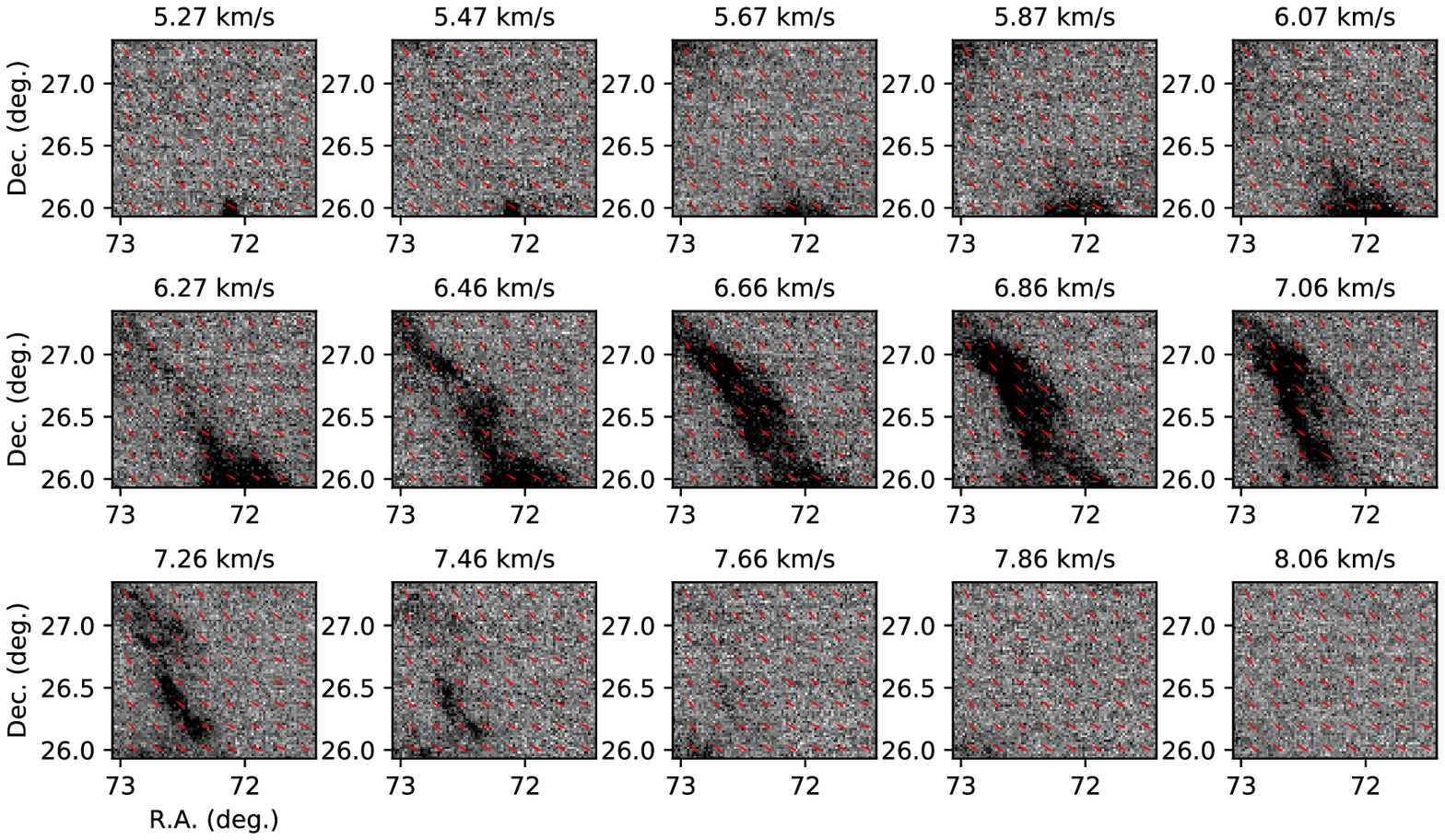,width=17cm,clip=}
\end{tabular}
\caption{A set of thin-slice velocity channel images of \co\ (top) and \coa\ (bottom) 
 J=1-0 emission from 1 subfield area (1.42$\times$1.42~deg$^2$)
in the vicinity of magnetically aligned striations in the Taurus molecular clouds \citep{Goldsmith:2008, Heyer:2016}.  
The red segments represent the 
orientation of the magnetic field derived from 353~GHz Planck polarization data with 10\arcmin\ resolution but displayed every 5.3\arcmin.   
The halftone of 
\co\ emission ranges from $-$1 to 2.5~K 
for each image.  For \coa\ emission, the halftone ranges from $-$0.5 to 0.5 K.
}
\label{figure2}
\end{figure*}
Figure~\ref{figure2} shows a mosaic of 15 thin-slice channel images of \co\ and \coa\ J=1-0 emission from a subfield in
the Taurus molecular cloud between \vlsr\ 5.24 and 8.06 \kms.  
To increase the signal to noise for each velocity slice image, the data are smoothed over 3 spectral channels providing a slice
thickness of 0.19 \kms\ for \co\ and 0.2 \kms\ for \coa.  
This velocity slice width is still smaller than the typical velocity 
dispersion of $\sim$1 \kms\  over the subfield size and therefore, 
within the thin-slice limit.
The red segments represent the 
orientations of the magnetic field component perpendicular to the plane of the sky 
 derived from 
the Planck 353~GHz polarization data. 
This particular field is the area noted by \citet{Goldsmith:2008} that exhibits emission striations and 
strong velocity anisotropy that is aligned with the magnetic field orientation \citep{Heyer:2008, Heyer:2012, Heyer:2016}.  
Therefore, it  
offers a best-case example of velocity anisotropy.
The emission from both isotopologues is weak (T$_{mb}$ $<$ 2~K) as this subfield is located in the 
periphery of the cloud that is well displaced from the 
high column density regions with active star formation.  The CO emission is stretched-out along the magnetic 
field orientation in many of these channel images.  Neighboring 
subfields exhibit similar alignment of the emission over these intervals.  

\subsection{Histogram of oriented gradients from the Taurus molecular data}
\begin{figure*}
\begin{tabular}{c}
\epsfig{file=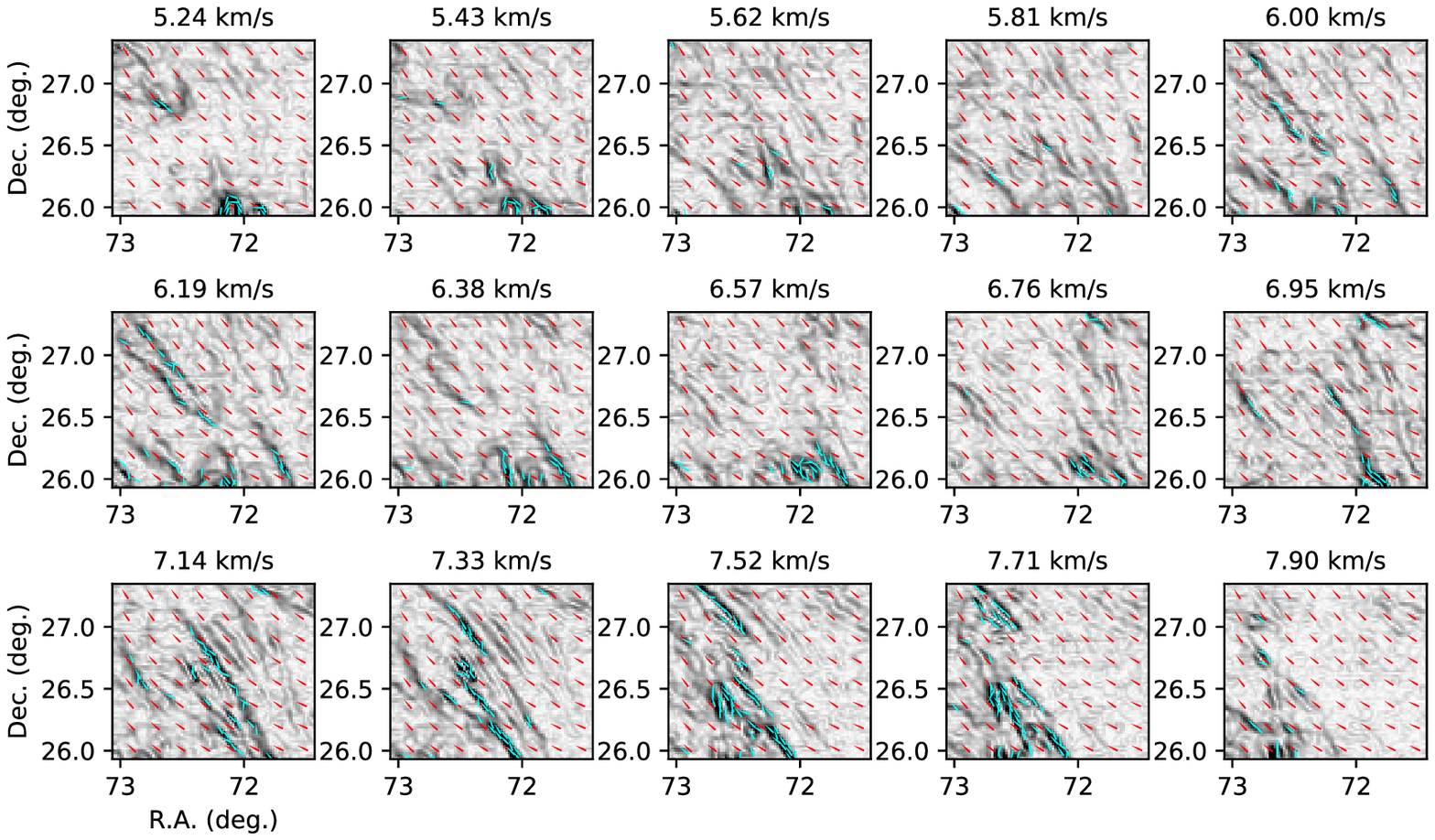,width=17cm,clip=}\\[0.5in]
\epsfig{file=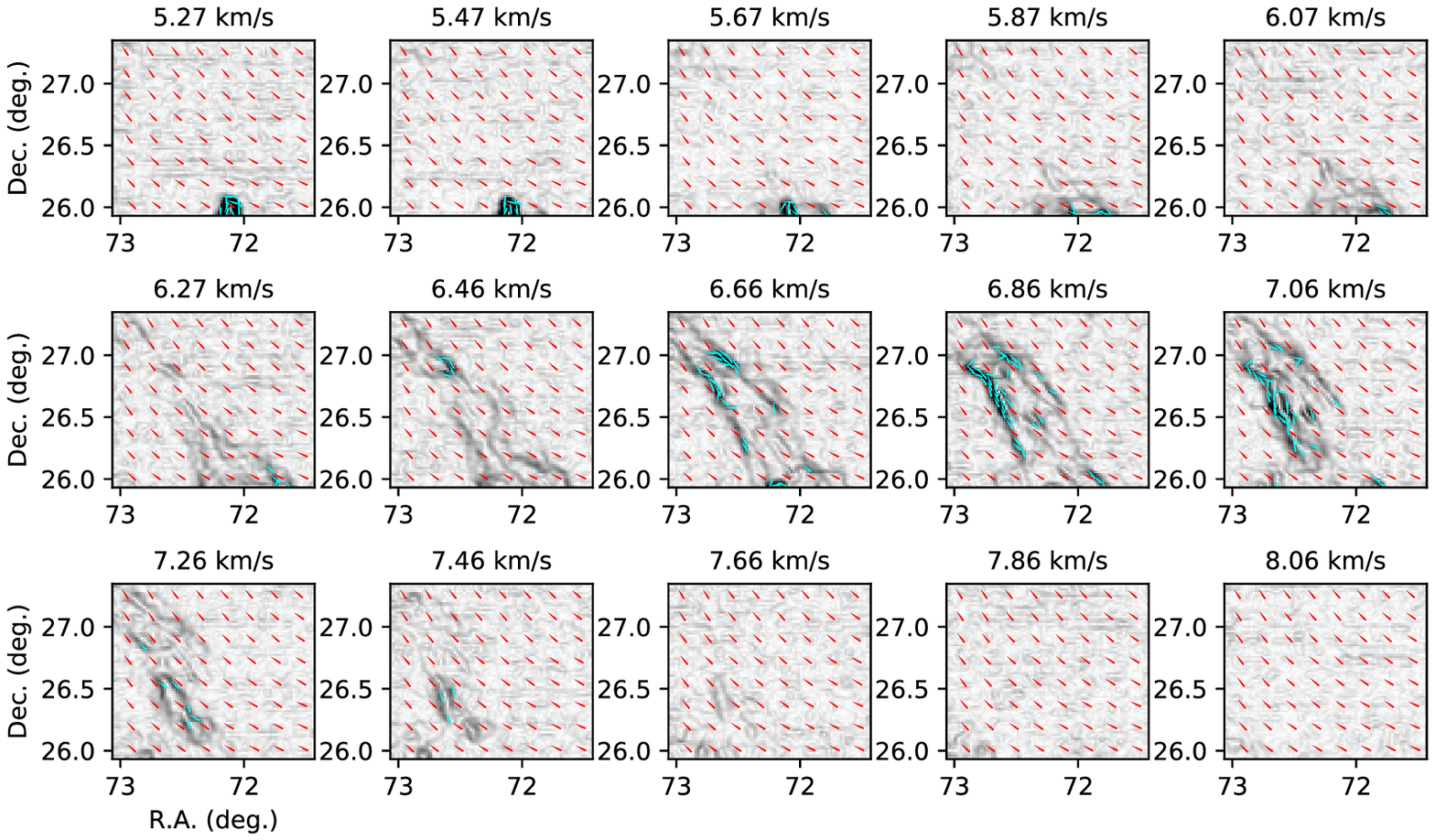,width=17cm,clip=}
\end{tabular}
\caption{ Images of gradient vector magnitudes of velocity slice images shown in Figure~\ref{figure2}. 
The halftone ranges from 0 to 8 km s$^{-1}$ pc$^{-1}$ for \co\ and 
0 to 4 km s$^{-1}$ pc$^{-1}$ for \coa.
The red segments show the orientation of the magnetic field component in the plane of the sky.
The cyan segments show the 
orientation of the gradient normal. 
Only segments with $|{\nabla}T(x,y,v_i)| >$ 4 K pc$^{-1}$ for \co\ and 
$|{\nabla}T(x,y,v_i)| >$ 2.2 K pc$^{-1}$ for \coa\ 
are shown for clarity. 
}
\label{figure3}
\end{figure*}

\begin{figure*}
\begin{tabular}{c}
\\
\\
\epsfig{file=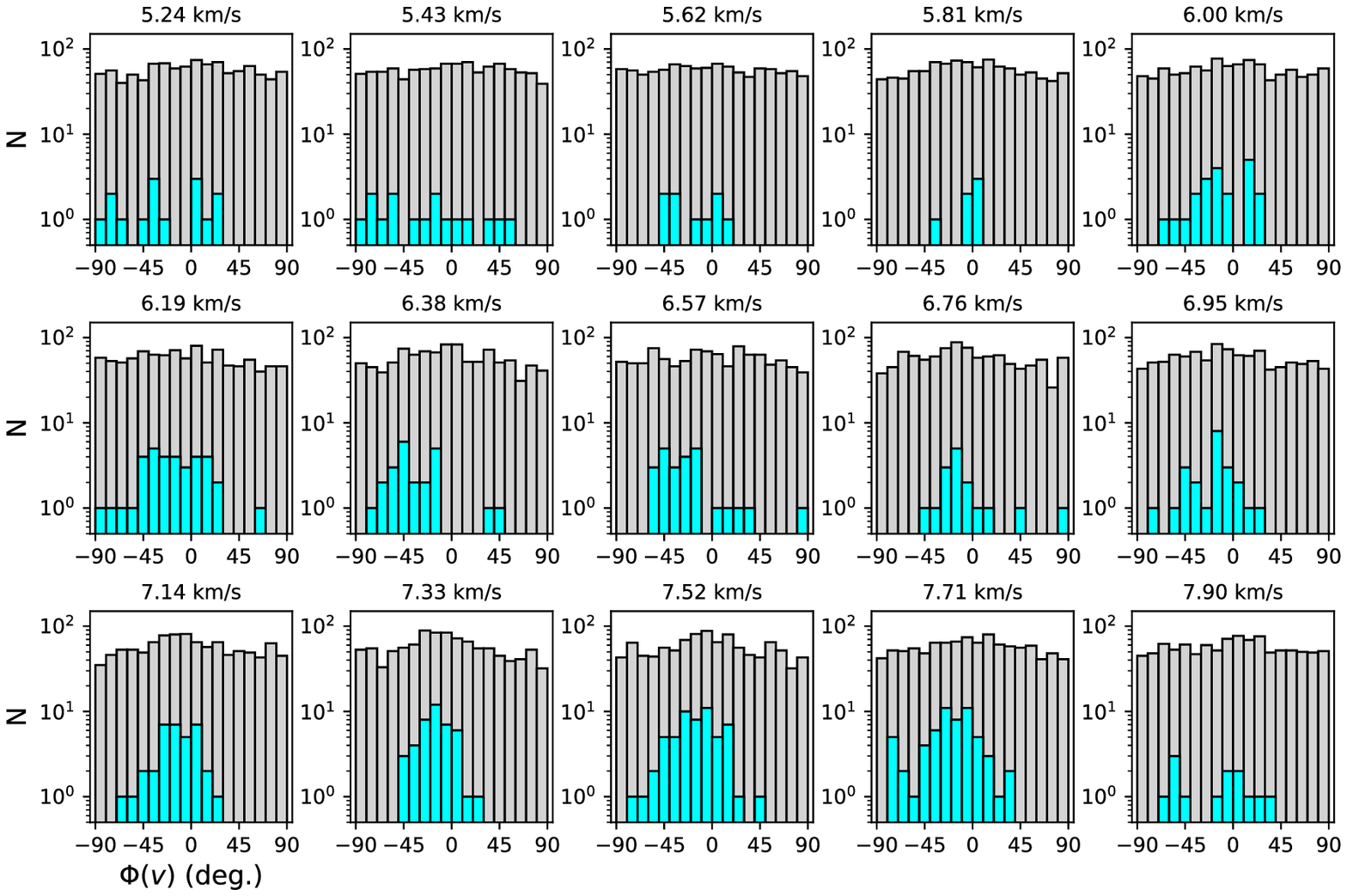,width=15cm,clip=}\\[0.5in]
\epsfig{file=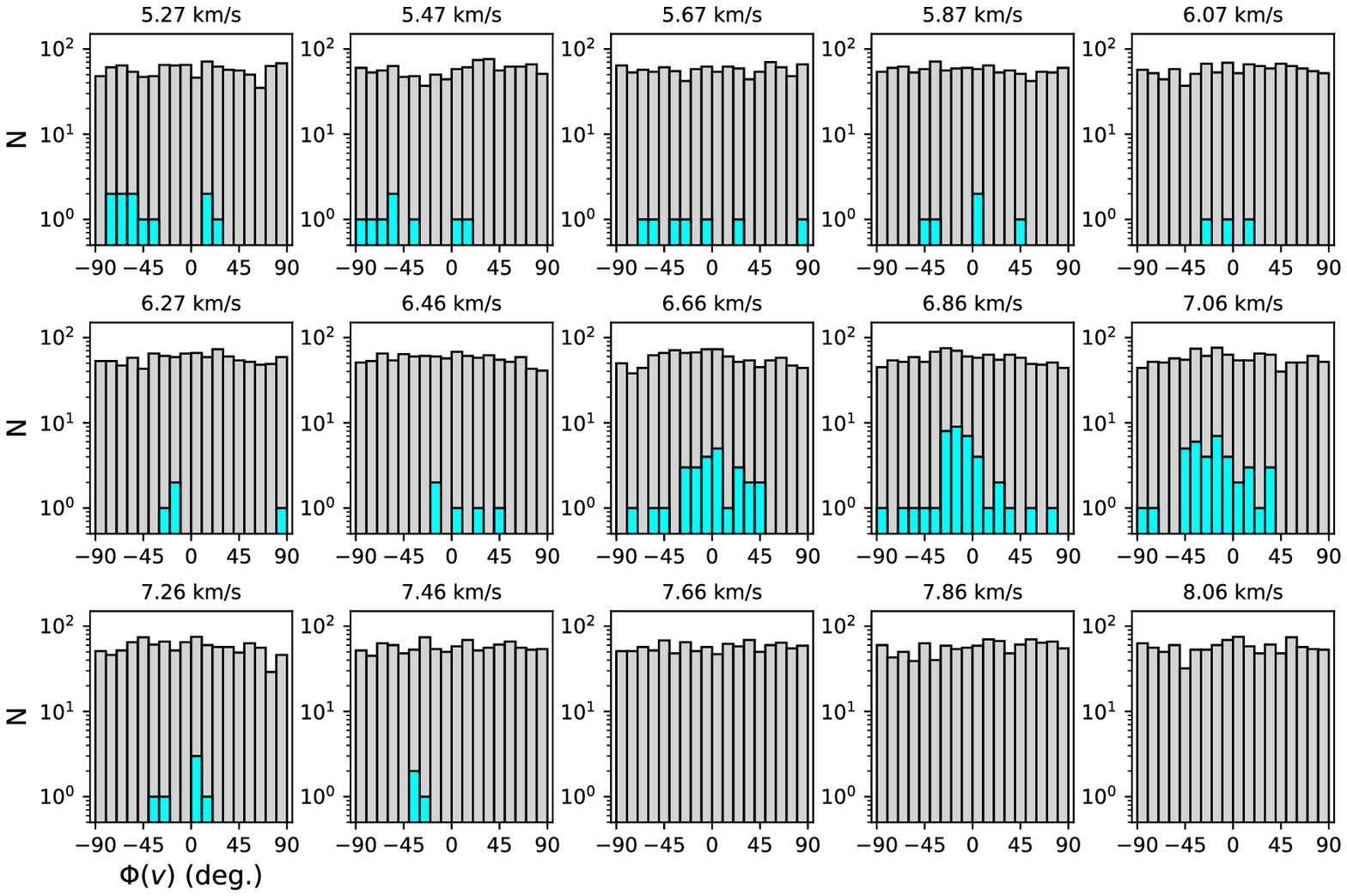,width=15cm,clip=}
\end{tabular}
\caption{ Histograms of the relative orientation between the gradient normal vector and the magnetic field orientation 
for each thin-slice field shown in Figure~\ref{figure2}. The light-grey shading shows the distribution of 
$\Phi$ values for all sampled pixels while the cyan histogram shows the distribution of 
$\Phi$ values for pixels with ${\nabla}I > $
4 K pc$^{-1}$ (\co) and 2.2 K pc$^{-1}$ (\coa).
}
\label{figure4}
\end{figure*}

The gradient analysis is applied to the 
set of \co\ and \coa\ 
velocity slices
between the V$_{LSR}$ interval -5 and 15 \kms\ for each subfield.  
Most of the signal from the Taurus cloud emits within \vlsr\ interval 4 to 8.5 \kms.  
The broader interval ensures sufficient baseline coverage to compile statistics of random noise. 
Figure~\ref{figure3} displays the image of gradient amplitudes 
derived from each velocity slice image shown in Figure~\ref{figure2}.
The cyan colored segments show those vectors with gradients that are greater than 4 K/pc (\co) and 2.2 K/pc (\coa).  These 
gradient thresholds are based on where the median error of the gradient angle is less than 15$^\circ$ (see Figure~\ref{figureA1}),  
and are only used 
to highlight the most statistically significant gradients.
The red segments represent $\Psi_B$.
The gradient features in Figure~\ref{figure3} typically persist for several contiguous channels (0.6 \kms).  
Many of the gradients in this field are aligned with 
the near uniform magnetic field orientations.   This alignment is particularly strong in channel images between 
6.9 and 7.5 \kms.  
The degree of alignment is further illustrated in Figure~\ref{figure4}, which shows histograms of 
the differences between the 
position angles of the gradient normal vector and the Planck-derived magnetic field orientations for each 
channel image in this demonstration field. 
The light grey histograms show the distributions of $\Phi_k$ values for all sampled pixels in each image while 
the cyan histograms show the $\Phi_k$ distribution for sampled pixels with ${\nabla}I >$ 4 K pc$^{-1}$ for \co\  
and ${\nabla}I >$ 2.2 K pc$^{-1}$ for \coa, 
for which the gradient angles are more reliably determined with respect to random noise effects.  
There are many positions in the field for a given velocity slice for which there is weak or no 
detected line emission.  The gradient angles from these pixels are random and produce a mostly flat 
$\Phi_k$ distribution. 
The cyan colored distributions in velocity channels 6.95 to 7.52 \kms\ are not precisely centered on zero degrees but 
show that the majority of pixels with well defined gradients have $|\Phi_k|$ values less than 20$^\circ$. These confirm the 
alignment that is 
visually evident in Figure~\ref{figure3}.  

\begin{figure}
\begin{center}
\epsfig{file=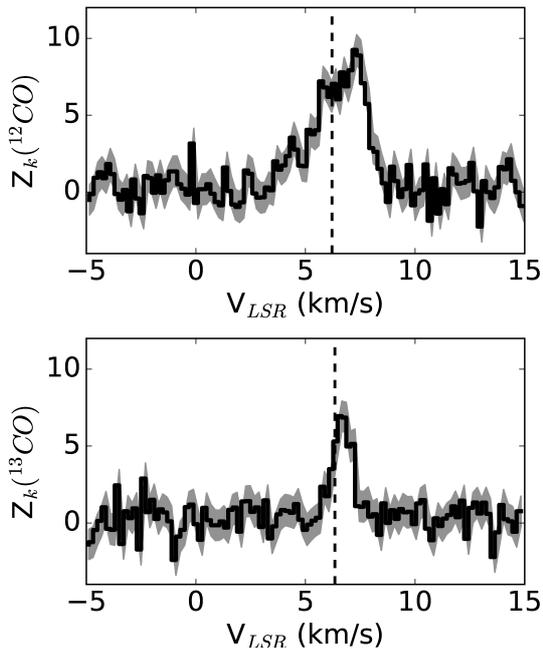,width=7cm,clip=}
\caption{ The variation of the projected Rayleigh statistic, $Z_k$  with \vlsr\ for the demonstration subfield shown 
in Figure~\ref{figure2} for \co\ (top) and \coa\ (bottom).  The light-grey shading shows the 1-$\sigma$ error in $Z_k$ values derived 
from Monte Carlo realizations. The vertical dashed line in each plot shows the mean velocity of the subfield 
field for each isotopologue. 
For both \co\ and \coa, the subfield shows statistically significant alignment of thin-slice intensity gradients with the magnetic field orientation in 
narrow velocity intervals. 
}
\label{figure5}
\end{center}
\end{figure}

The projected Rayleigh statistic, $Z_{k}$, described in \S3.3 offers a more quantitative assessment of the degree 
of alignment for a set of relative orientations than visual inspections of histograms. 
The variations of $Z_{k}$ values for each computed velocity slice for the demonstration field for 
both \co\ and \coa\ are presented in Figure~\ref{figure5}. 
In calculating $Z_{k}$ for a given image, all sampled pixels are used 
but are weighted by the factor $1/\sigma(\Phi_{k})^2$, where $\sigma(\Phi_{k})$ is the random 
error in $\Phi_{k}$ values derived for each sampled pixel 
as estimated from the Monte Carlo calculation of $\Psi_{N,k}$ as well as uncertainties of $U$ and $Q$ values. 
Examining Figure~\ref{figure5}, it is useful to first inspect values in the velocity 
intervals with no emission.  For this field, these intervals are [-5,+2] \kms\ and [10,15] \kms.  Ideally, the random noise in 
these images should generate $Z_{k}$ values near zero. The small but visible  bias of $Z_k$ towards small negative values 
is likely a result of spatially correlated noise of the spectral line data that arises from OTF mapping.   As already 
described, the standard deviation of $Z_k$ values in these intervals is $\sim$1. 
For this field, most of the emission occurs within the velocity interval [4.5,9] \kms (see Figure~1).  Within 
this interval, the $Z_k$ values are positive and often greater than 3 corresponding to signal to noise greater than 3 and 
indicative of mostly parallal alignments.
The largest $Z_k$ values occur within channels that are slightly red-shifted with respect to the mean velocity of this subfield 
-- a feature 
previously noted by \citet{Heyer:2016} in this field. 

\subsection{All sub-fields} 
Figures 2-5 provide a demonstration of the gradient analysis for a single subfield. We have applied this method to all 150 subfields
to assess the degree of gradient alignment throughout the Taurus cloud and its dependence on local gas properties. 
Previous studies have shown that the magnetic field orientations derived from the Planck polarization data vary smoothly across the 
Taurus cloud 
\citep{Planck:2015_XXXV, Hu:2019}.  At 353~GHz,
the thermal dust emission is optically thin so these orientations should represent the density weighted average of the magnetic field orientation along the 
line of sight.  However, we caution that this projected magnetic field orientation may not reflect the 3-dimensional topology of the magnetic field within 
the gas layers traced by each
thin velocity slice. Any decoupling of gas layers respectively traced by the polarization and thin-slice line 
emission would introduce increased variance in the distribution of $\Phi_k$. 

\begin{figure*}
\begin{tabular}{c}
\\
\\
\epsfig{file=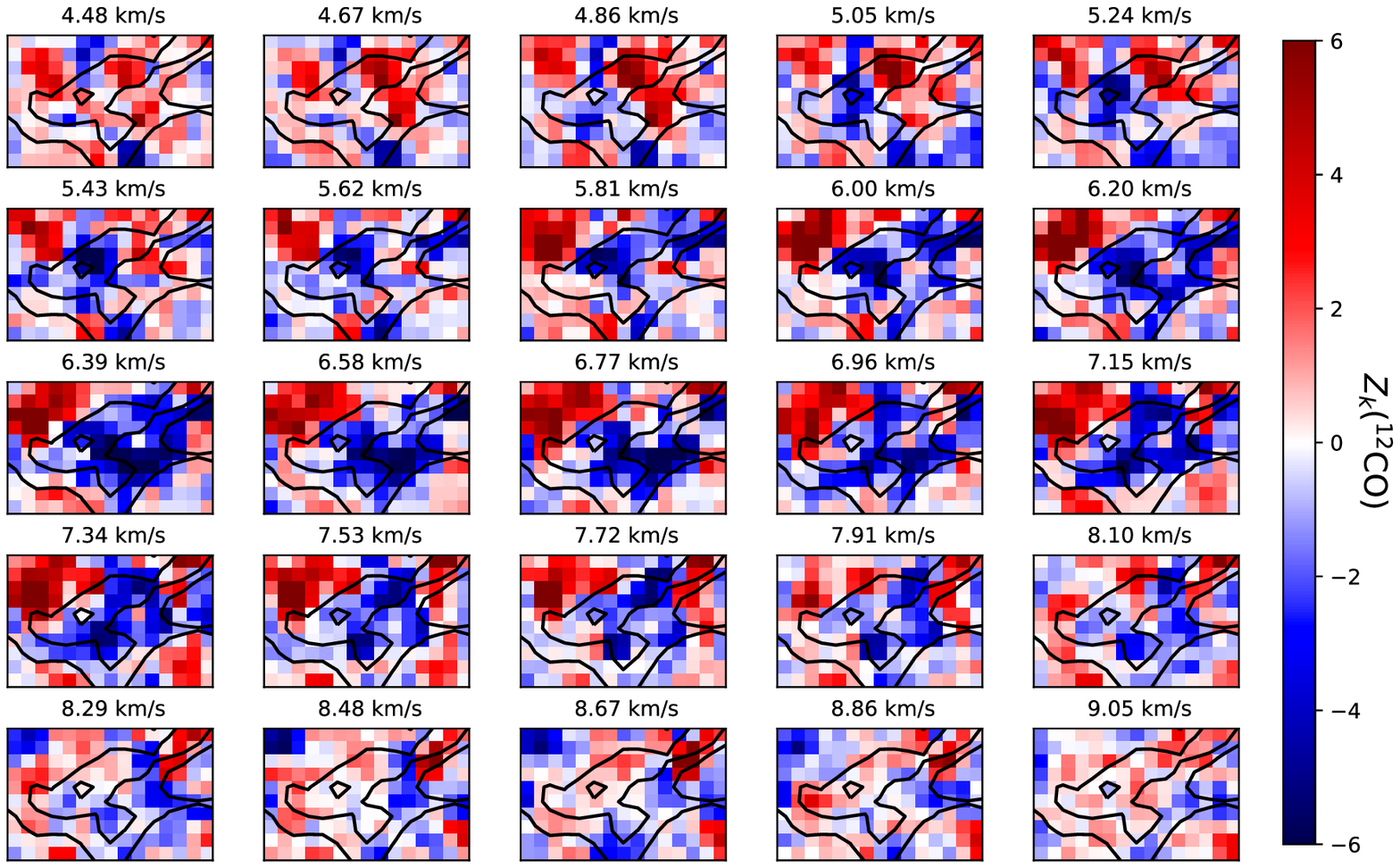,width=15cm,clip=}\\[0.5in]
\epsfig{file=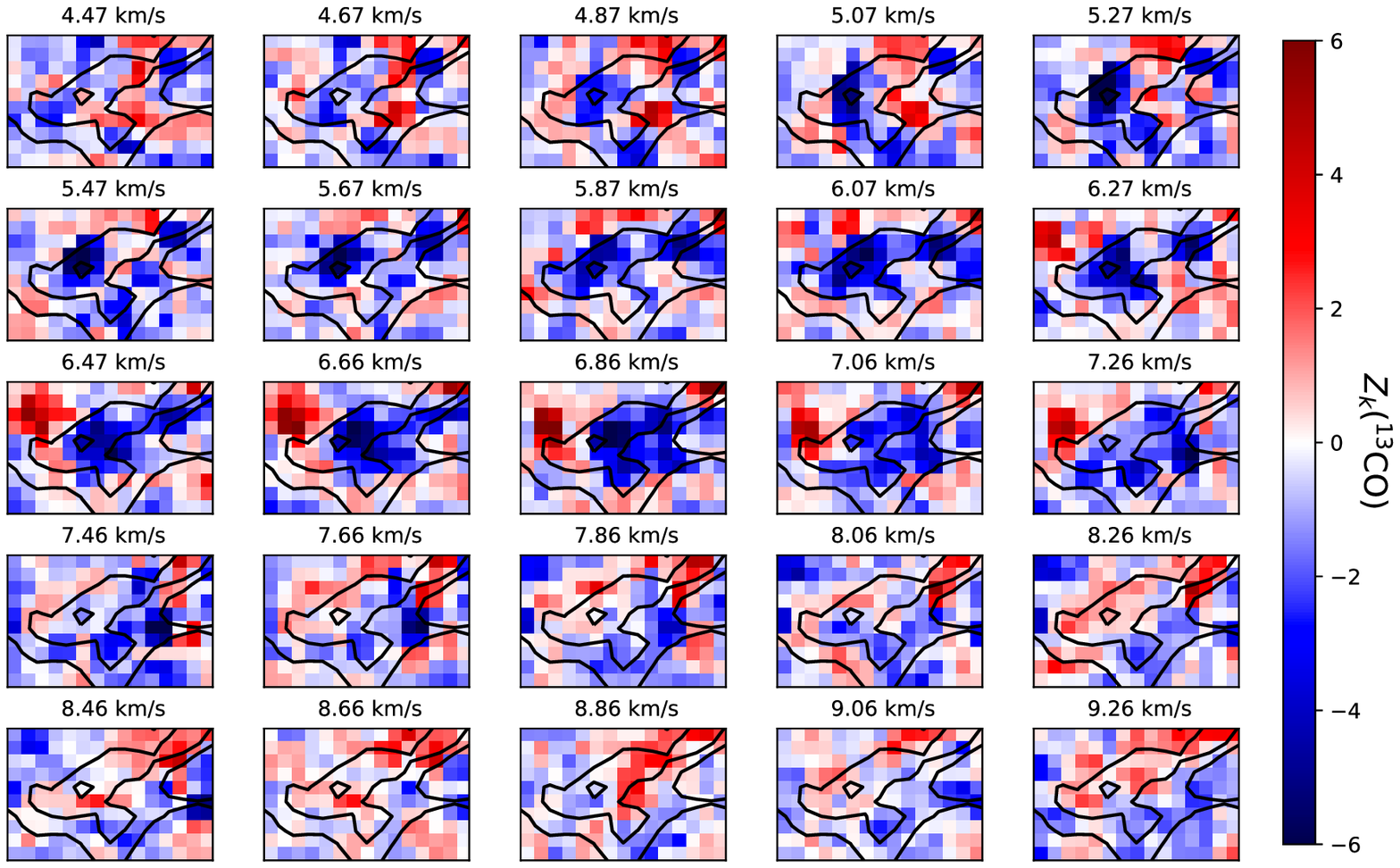,width=15cm,clip=}
\end{tabular}
\caption{ Images of the projected Rayleigh statisitic, $Z_k$, for \co\ (top) and \coa\ (bottom)
calculated over each subfield of the Taurus molecular cloud. Axis labels are not shown for clarity but would be 
identical to those in Figure~\ref{figure7}. 
The contours represent the log of the hydrogen column density (cm$^{-2}$) derived from Planck dust emission that has 
been resampled to the effective 
resolution of the $Z_k$ image.  The contour levels are {21.5, 21.75, and 22}. 
There is strong alignment of the velocity gradient normal vector position angle with the magnetic field orientation (red colors) 
in the northeast for most thin-slice images.  In the highest column density regions, there is a strong signal of perpendicular 
relative orientation (blue colors) in the \coa\ data that more accurately traces such regions relative to \co.
}
\label{figure6}
\end{figure*}

Figure~\ref{figure6} shows images of $Z_k$ values derived from the \co\ and \coa\ data for a subset of all analyzed 
velocity slices. Each pixel in these maps represents the $Z_k$ value for each subfield.  The color lookup table is diverging in order to 
identify subfields with thin-slice intensity gradients that are strongly aligned (red), perpendicular (blue), or random (white). 
The overlayed contours show the distribution of hydrogen column density from the Planck data but sampled at the resolution of 
each subfield (1.42$^\circ$ or 3.5~pc).  The 1$\sigma$ uncertainty in $Z_k$ values is $\sim$1 so only the strongly 
red or blue colors denote significant 
alignment (parallel or perpendicular).  In both the \co\ and \coa\ data, 
the northeast region of the cloud exhibits the largest positive $Z_k$ values
over the interval 5.75 to 7.75 \kms. These striation features are elongated 
along the local magnetic field direction.   For 
the high column density regions, 
which typically correspond to dense filaments in both dust continuum and \coa\ 
line emissions, 
$Z_k$ values are mostly negative over the velocity 
range 5 to 7 \kms.  These negative values are a result of the 
edges of the filaments, which generate large gradients,  
whose normal orientations are perpendicular to the local magnetic field. 

\subsection{Projected Velocity integrated Rayleigh Statistic}
Magnetically aligned anisotropies are 
 evident in the gradient images over very narrow velocity intervals -- 
typically less than 1 \kms \citep{Heyer:2016}.  Here, we address whether 
alignments are coherent over larger velocity intervals as inferred in Figure~\ref{figure5}.  The 
velocity-integrated Rayleigh statistic is 
\begin{equation}
Z_{int} = \frac{\sum_k \sum_i w_k(x_i,y_i) \cos(2\Phi_{k}(x_i,y_i)) } { (\sum_k \sum_i w_{k}(x_i,y_i)^2)^{1/2} / 2}
\end{equation}
where the sum i is over all sampled pixels in each subfield and the sum k is over all velocities within the interval 5 to 8 \kms.  This velocity 
interval accounts for 69\% and 83\% of the \co\ and \coa\ luminosities respectively.  
$Z_{int}$ is similar to the method of \citet{Hu:2019}, which analyzes 
the histogram of all thin-slice gradient orientations compiled over a larger velocity interval 
to derive the relative orientation of a subfield. 

The projected velocity integrated Rayleigh statistic is complemented by the velocity-integrated mean resultant vector,
\begin{equation}
r_{int} = \frac{\Bigl[C^2+S^2\Bigr]^{1/2}}{\Bigl(\sum_k \sum_i w_{k}(x_i,y_i) \Bigr)}
\end{equation}
where
\begin{equation}
C = \sum_k \sum_i w_k(x_i,y_i) \cos(2\Phi_{k}(x_i,y_i)) 
\end{equation}
\begin{equation}
S = \sum_k \sum_i w_k(x_i,y_i) \sin(2\Phi_{k}(x_i,y_i)) 
\end{equation}
The mean resultant vector is the quadrature sum of 2 terms: $C$ measures whether there is a preferred 
orientation with respect to 0 or 90 degrees; $S$ measures whether there is a preferred orientation with respect to 45 or 135 degrees.
Unweighted, it is the fraction of gradient 
vectors pointing in a preferred direction but does not carry information on what that preferred direction is \citep{Soler:2019}. 
Weighting the calculation by $1/\sigma(\Phi_k(x_i,y_i))^2$, $r_{int}$ reflects 
this fraction biased towards pixels with small 
instrumental $\Phi$ errors and provides a normalized estimate of the 
significance of the alignment.  The values of $Z_{int}$ and $r_{int}$ are 
used together to establish the degree of alignment between two orientations.  The mean resultant vector also resolves
 the degeneracy of 
$Z$ for a set of random orientations for which $Z$=0 and $r$=0 and a set of orientations all at 45$^\circ$ for which $Z$=0 and $r$=1.

\begin{figure*}
\begin{tabular}{c}
\epsfig{file=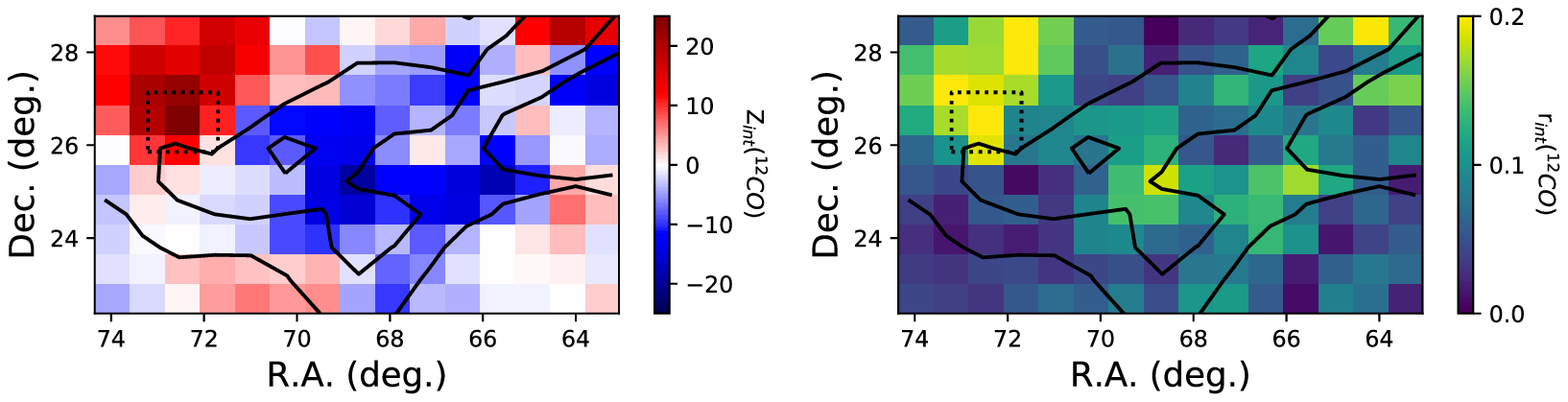,width=17cm,clip=}\\[0.25in]
\epsfig{file=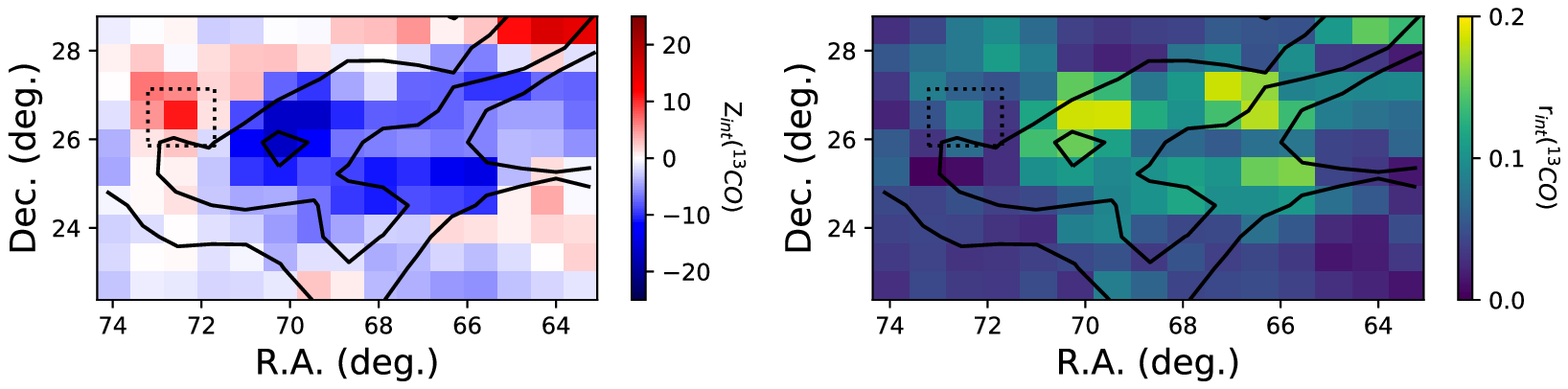,width=17cm,clip=}
\end{tabular}
\caption{Images of the projected Rayleigh statistic, (left) $Z_{int}$ and (right) mean resultant vector, $r_{int}$ 
for \co\ (top) and \coa\ (bottom) 
integrated over velocities 5 to 8 \kms\  for each subfield.  The dotted-line box in each panel 
shows the location and area of the demonstration field presented in Figure~\ref{figure2}.
The contours are identical to those in Figure~\ref{figure6}. 
}
\label{figure7}
\end{figure*}

Images of $Z_{int}$ and $r_{int}$ for \co\ and \coa\ are shown in Figure~\ref{figure7}.  Submaps with higher values of $r_{int}$ 
have more reliable measures of $Z_{int}$ values.  As shown in Figure~\ref{figureB1} in the Appendix, 
values of $r_{int} > 0.1$ 
correspond to $Z_{int}/\sigma(Z_{int}) > 3$.
The range of $Z_{int}$ values is 3 times larger than that of $Z_k$ values for thin-slice channels showing 
that magnetic alignments of gradient normals  accumulate over 
larger velocity intervals.  In general, there is good correspondance between $Z_k$ and $Z_{int}$ for both \co\ and \coa. 
The same subfields where $Z_k > 3$ in the thin-slice channels 
of Figure~\ref{figure6} 
are also locations where $Z_{int} > 3$.  Both isotopologues show mostly 
perpendicular orientations in the central region of the cloud. 
Quantitatively, for \co, 28\% of the subfields have $Z_{int}$($^{12}$CO) $ >$ 3 (mostly parallel), 39\% of the subfields have
$Z_{int}$($^{12}$CO) $ <$ -3 (mostly perpendicular), and 33\% have $|Z_{int}$($^{12}$CO)| $<$ 3 (no preferred direction).
For \coa, these percentages are 7\%, 43\%, and 50\%. 
The strongly parallel or perpendicular subfields are localized to specific areas in the cloud.  Much of the cloud area exhibits 
no preferred alignment either due to an absence of signal or a broad distribution of $\Phi_k$ values 
derived from significant gradients.

\section{Variation of relative orientations  with gas properties}
In the previous section, we find a broad range of relative orientations between thin-slice 
intensity gradients and the magnetic field. This diversity of orientations is not exclusively due to 
random errors but likely reflects spatial variations of conditions such as the magnetization, 
ion-neutral coupling, gas to magnetic flux ratios, efficiency of alignment of dust grains, 
or the projection of 3-dimensional quantities into 
the two-dimensional frame of the observations \citep{Seifried:2020}. 
In this section, we examine the conditions in the cloud accessible from the \co, \coa, and Planck  
data that may regulate 
the alignment of elongated turbulent eddies or gas flows with the
magnetic field orientation.

\subsection{Column Density}
The neutral gas component of clouds is coupled to the interstellar magnetic field through collisions 
with ions.  The ionization fraction of the gas is maintained by cosmic rays and 
the ultraviolet radiation field.  
In the low column density envelopes of molecular clouds, there is more exposure to ambient 
photoionizing radiation that
can keep the neutral gas well coupled to the interstellar magnetic field.  With a stronger coupling of ions and 
neutrals, it 
is reasonable to expect a stronger alignment of 
MHD turbulence or large scale flows with the magnetic field orientation 
in regions of low column density.


Figure~\ref{figure6} and Figure~\ref{figure7} show a stronger 
degree of alignment (large $Z_k$ and $Z_{int}$ values) of the vector normal to the thin-slice intensity gradient vector 
with the magnetic field in the outer edges of the \co\ and \coa\ maps.
This alignment is particularly strong in the northeast sector of the cloud
in both \co\ and \coa, where the magnetically aligned striations are found \citep{Goldsmith:2008, Heyer:2016} and 
where PCA-derived velocity structure functions are anisotropic and aligned with the magnetic field orientation \citep{Heyer:2008, Heyer:2012}. 
This region is characterized by low column densities inferred from visual extinction maps \citep{Pineda:2010} and Planck-derived opacity images
 \citep{Planck:2015_XXXV}
and low density, sub-thermal  excitation conditions of \co\ and \coa\ J=1-0 lines \citep{Pineda:2010}.  
Conversely, in the central regions with higher column densities, there is a preferred perpendicular alignment. 

\begin{figure*}
\epsfig{file=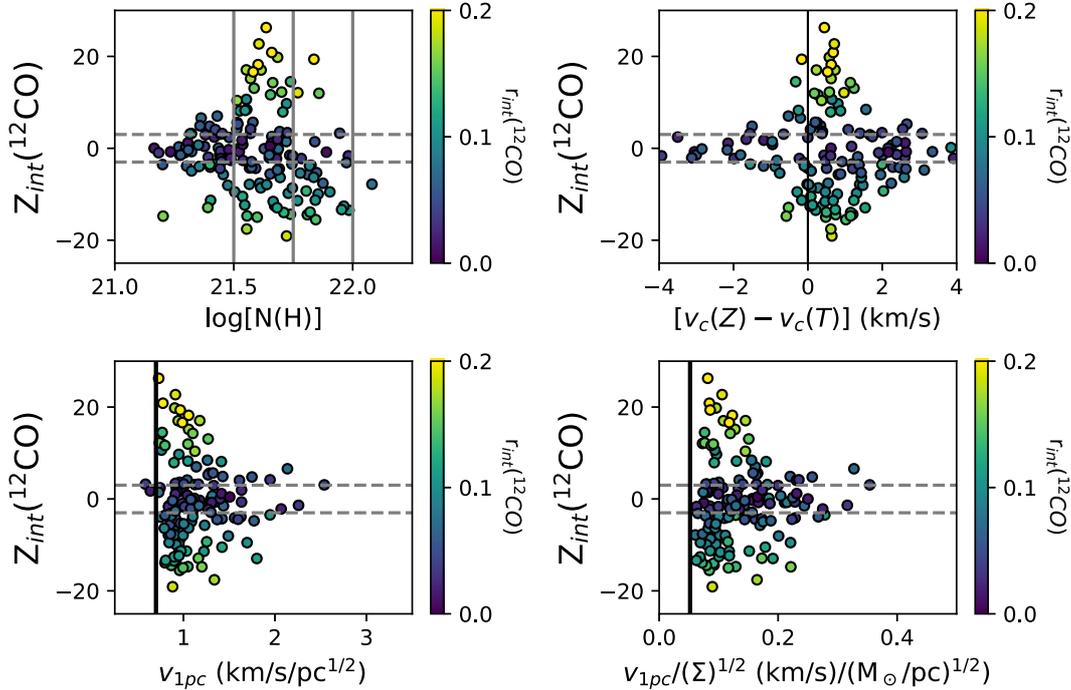,width=14cm,clip=}
\caption{Variation of $Z_{int}$($^{12}$CO)  values in each subfield with 
(top left) hydrogen column density derived from Planck and smoothed to subfield angular 
resolution with vertical gray lines marking the contour levels in Figure~\ref{figure6} and Figure~\ref{figure7}, 
(top right) offset between first moments of $Z_k$ and $T_k$ with vertical line denoting zero offset, 
(bottom left) v$_{1pc}$, and (bottom right)  v$_{1pc}/\Sigma^{1/2}$. 
The color of each point represents the value of the mean resultant vector with 
yellow-green-teal points more reliable than blue-purple points.  The horizontal dashed lines represent 
$|Z_{int}|/\sigma(Z_{int})=3$.  
The vertical solid lines in the bottom panels represent the mean values of v$_{1pc}$ and v$_{1pc}/\Sigma^{1/2}$
derived for Galactic 
clouds. 
}
\label{figure8}
\end{figure*}

\begin{figure*}
\epsfig{file=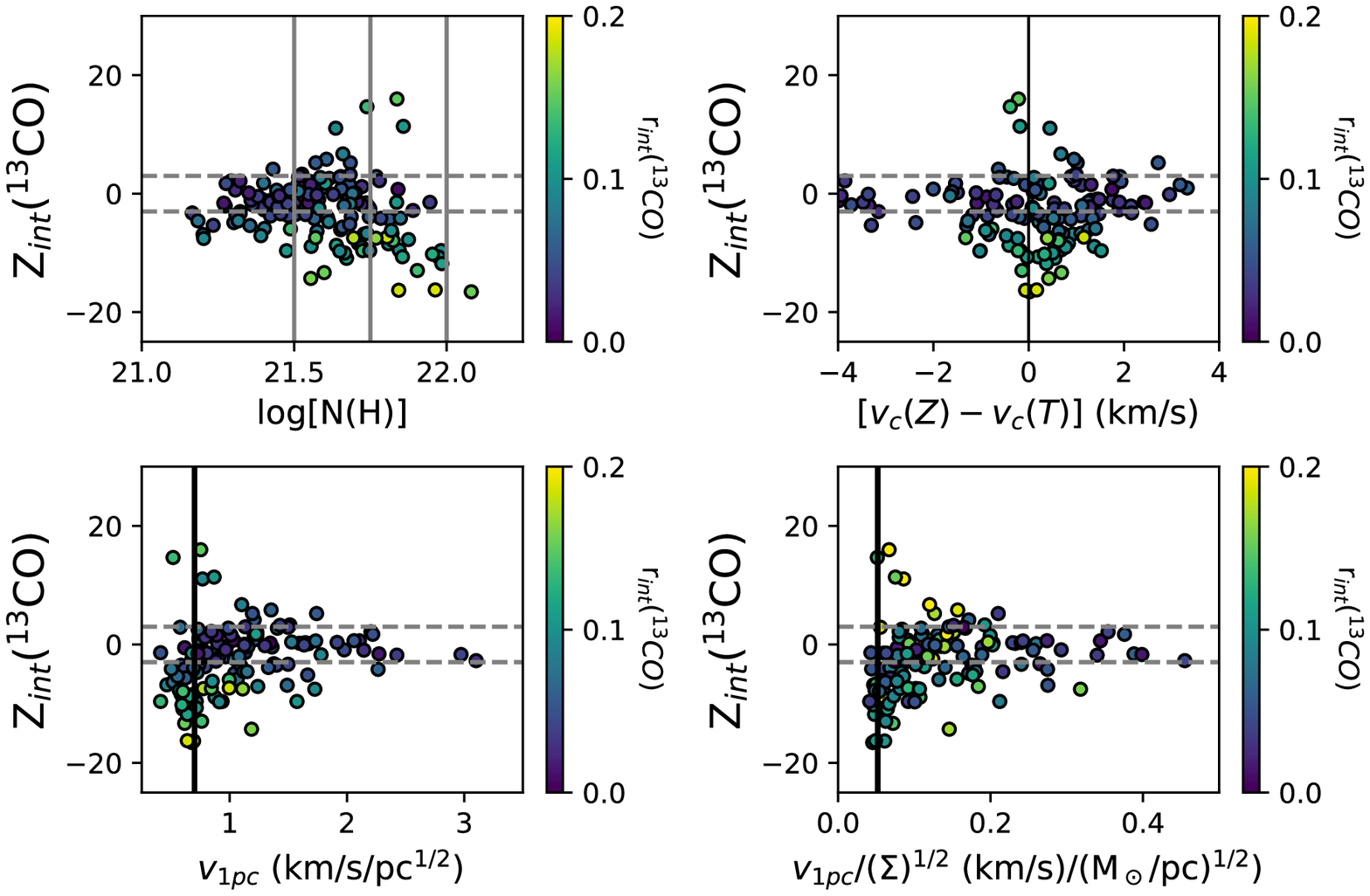,width=14cm,clip=}
\caption{Same as Figure~\ref{figure8} for \coa. 
}
\label{figure9}
\end{figure*}

Does this trend of relative orientations with column density hold throughout the cloud?
To examine this question, we plot the variation of $Z_{int}$ values 
with column density 
over each subfield area in the upper left boxes of Figure~\ref{figure8} and Figure~\ref{figure9}.  
The mean resultant vector for each subfield 
is encapsulated in 
the color of each point.  
To guide the eye, the vertical gray lines correspond to the drawn contour 
levels in Figure~\ref{figure6} and Figure~\ref{figure7}. 
For both \co\ and \coa\, there are significant positive and negative values of $Z_{int}$
over the log(N(H)) range 21.5-21.75.  Evidently, column density is not the only factor that 
determines whether turbulent eddies are aligned along the magnetic field.

At higher column densities, where one must 
be more cautious in interpreting the \co\ alignments owing to large line opacities, 
there is a clear excess of subfields with negative $Z_{int}$($^{12}$CO) values for
log(N(H)) $>$ 21.75.  This excess is also present when evaluated at higher resolutions (smaller
subfield sizes) as shown in Figure~\ref{figureC1}. 
The degree of alignments in this high column density regime is best probed with lower 
opacity \coa\ data (see Figure~\ref{figure9}).  Again, for log(N(H)) $>$ 21.75, 
there many more subfields with $Z_{int}$($^{13}$CO) values less than $-$3 corresponding to 
perpendicular alignment than those with values greater than 3 (see also Figure~\ref{figureC2}). 

\subsection{Velocity Displacement}
In the demonstration field, we found that the peak value of the $Z_k$ spectrum 
 is offset from the 
velocity centroid of the average \co\ and \coa\ spectra over the subfield (see Figure~\ref{figure5}).  
One possible reason for 
this displacement is reduced confusion along the line of sight as one is more kinematically displaced 
from the bulk of the emission or simply lower opacity in the line wings.  To assess whether this displacement is 
a factor in aligning the 
thin-slice intensity gradient with the magnetic field orientation, we plot the variation of $Z_k$ with 
the difference of velocity centroids respectively calculated from $Z_k$ and the average brightness temperature 
spectrum, T(v), for each subfield.  The results are shown in the upper right boxes of Figure~\ref{figure8} and Figure~\ref{figure9}.
For \co, this displacement is $\sim$ 0.5 \kms (redshifted) for both positive and negative values of $Z_{int}$.
For \coa\, the distribution of velocity displacements of significant points ($r_{int}$ $>$ 0.1) is more centered on zero. 
This difference between 
\co\ and \coa\ velocity displacements suggests that observed alignments are limited to lines of sight or velocity displacements 
where the line opacity is low. 
For \coa, this condition is met at line center owing to its smaller abundance relative to \co.  
For \co, reduced optical depths over small velocity intervals, but still greater than unity, are possible in 
the wings of the spectral line profile. It is not clear why these displacements for \co\ 
are strongly skewed towards the redshifted side of the 
mean spectrum.  We speculate that blue shifted material may be located on the backside of the cloud so any emission from this layer 
would cross multiple correlation lengths that may dilute any signature of alignment. 

\subsection{Normalizations of Turbulence}
The alignment of turbulent eddies with the local magnetic field is a fundamental prediction of sub-Alfv\'enic 
MHD turbulence theories \citep{Goldreich:1995, Lazarian:1999, Cho:2003}.  Such alignment is also 
expected for magnetosonic waves \citep{Mouschovias:2011}. 
Moreover, our analysis of thin-slice intensities of molecular line emission as a measure of the turbulent velocity field is based on \citet{Lazarian:2000}
who analytically demonstrated that intensity spatial variations in the thin-slice limit arise primarily from velocity fluctuations. 
A useful measure to characterize the degree of turbulence within an area is the velocity dispersion measured at 
1~pc scale.  This corresponds to the 
normalization coefficient of the first-order structure function but can be estimated by simply evaluating the velocity dispersion, $\sigma_v(L)$ 
over a scale, $L$ in pc,
such that v$_{1pc}=\sigma_v(L/2)/(L/2)^{1/2}$, which assumes the power law index of the structure function is 1/2. 
The cloud to cloud velocity dispersion-size relationship emerges from the near invariance of this value when 
integrated over the whole cloud \citep{Larson:1981, Heyer:2004}. But 
this value can fluctuate within a cloud depending on local feedback processes from star formation
or 
the dependency of this parameter 
with surface density \citep{Heyer:2009}.  
A secondary measure of the turbulent motions is the quantity v$_{1pc}$/$\Sigma^{1/2}$, where $\Sigma$ is the mass surface density in units M$_\odot$/pc$^2$.

In the bottom rows of Figure~\ref{figure8} and Figure~\ref{figure9}, we show the variation of $Z_{int}$ values with 
v$_{1pc}=\sigma_v/(L/2)^{1/2}$  and v$_{1pc}/\Sigma^{1/2}$ for each subfield.  The velocity dispersions are 
derived from the second moments of the mean \co\ and \coa\ spectra constructed from 
all spectra within a given subfield.  Since all subfields have 
an identical size, 3.5~pc or $L/2$=1.75~pc, the dependency rests exclusively on the velocity dispersion.  
However, by casting the dependence as v$_{1pc}$, 
we can compare the values to those derived from more established cloud to cloud normalizations as shown by the vertical 
solid line.  The mass surface densities for each subfield are taken from the smoothed and aligned Planck column density 
image. 

The median values of v$_{1pc}$ 
are 1.0 and 0.9 for \co\ and \coa\ respectively with standard deviations of 0.4 and 0.7. 
The median values of v$_{1pc}$/$\Sigma^{1/2}$
are 0.12 and 0.10 for \co\ and \coa\ respectively with standard deviations of 0.07 and 0.11. 
For both parameters, these median values are slightly larger than the values
 derived for Galactic clouds by \citet{Solomon:1987, Heyer:2009}. 
There is no strong correlation of $Z_{int}$ with v$_{1pc}$ or v$_{1pc}/\Sigma^{1/2}$ in these plots.
However, we note that 
the most extreme values of $Z_{int}(^{12}{\rm CO})$ and $Z_{int}(^{13}{\rm CO}$) have 
values of v$_{1pc}$ and v$_{1pc}/\Sigma^{1/2}$ close to that of Galactic clouds.  Conversely, subfields with 
large values of these quantities have near zero $Z_{int}$ values corresponding to random orientations between the 
direction of intensity gradient normals and the magnetic field.  
The trend of strongly aligned gradients with Galactic values of v$_{1pc}$ and 
v$_{1pc}$/$\Sigma^{1/2}$ suggests that alignments occur only in turbulent-quiet 
regions. Given the MHD simulation results that anisotropy is limited to regions 
with sub-Alfv\'enic 
motions, we speculate that such quiet regions also correspond to 
sub-Alfv\'enic conditions. 

\subsection{Line Opacity}
The differential views of the Taurus cloud provided by \co\ and \coa\ J=1-0 emission 
as shown in Figure~\ref{figure1} are a result of very different line opacities.  The \co\ line 
is almost always optically thick in most conditions of molecular clouds where the \coa\ line is 
optically thin for most lines of sight within a cloud.  
In this section we examine the relationships between gradient alignments, velocity displacement 
from the line center of the average spectrum of a subfield, ${\rm v}_k-{\rm v}_c$, and $R_k$,  the 
ratio of \coa\ to \co\ antenna temperatures in 
velocity 
slice $k$, for each subfield.  
$R_k$ provides an observational proxy to the \co\ optical depth of emission within a velocity channel.  
\begin{equation}
R_k = \frac{\nu(^{13}{\rm CO})}{\nu(^{12}{\rm CO})}\; \frac{[J_\nu(T_x(^{13}{\rm CO}))-J_\nu(2.7)]}{[J_\nu(T_x(^{12}{\rm CO}))-J_\nu(2.7)]}
\frac{[1-e^{-\tau_k(^{13}{\rm CO})}]}{[1-e^{-\tau_k(^{12}{\rm CO})}]}
\end{equation}
where $\nu$ is the line rest frequency, $J_\nu(T)=1/(e^{h\nu/KT}-1)$, $T_x$ is the excitation temperature, and 
$\tau_k$ is the optical depth for channel $k$.  By assuming $\tau_{k}(^{13}{\rm CO})=\tau_{k}(^{12}{\rm CO})/X_{iso}$, 
where $X_{iso}$ is the abundance ratio of \co\ to \coa,
and values for the excitation temperatures, one can estimate $\tau_{k}(^{12}{\rm CO})$ and $\tau_{k}(^{13}{\rm CO})$. 
$R_k$ increases with increasing 
optical depth for a given 
set of values of $X_{iso}$, $T_x(^{12}{\rm CO})$, and $T_x(^{13}{\rm CO})$.

\begin{figure*}
\begin{center}
\epsfig{file=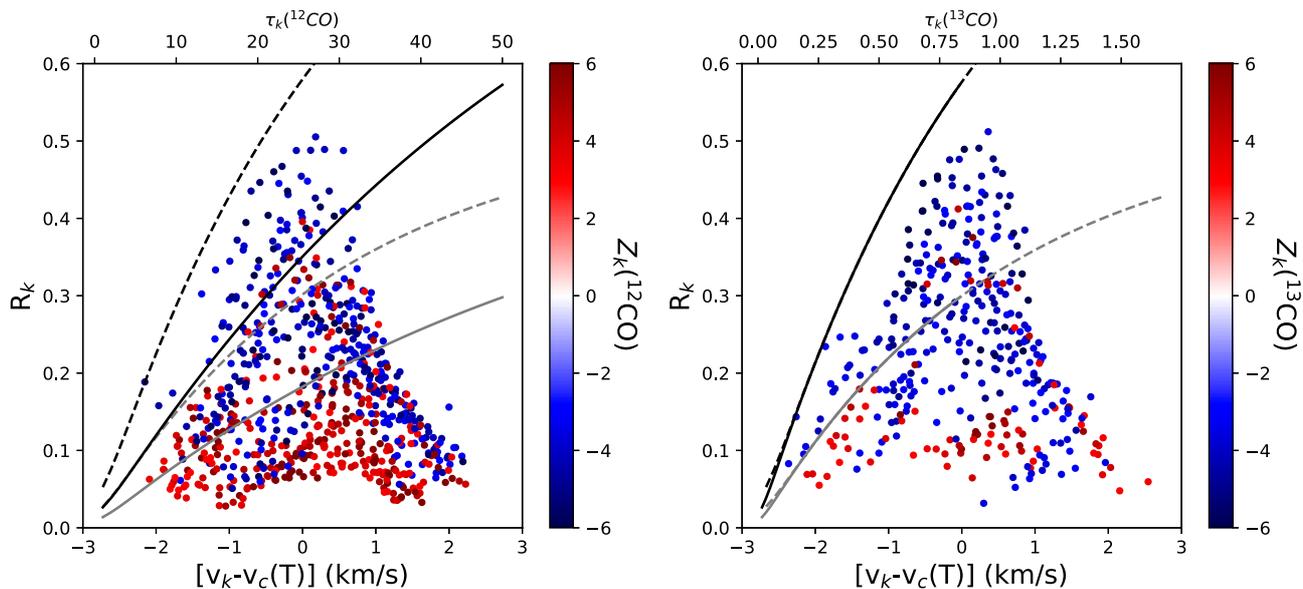,width=17cm,clip=}
\caption{Plots of $R_k \equiv \left< T_{k}(^{13}{\rm CO}) \right>/\left<T_{k}(^{12}{\rm CO})\right>$
values with velocity displacement for each subfield.  Point colors represent $Z_k$ values indicating parallel
(red) and perpendicular (blue) alignments for \co (left) and \coa (right).  
Only subfields and velocity slices with $R_k/\sigma(R_k) > 3$ 
and $Z_k/\sigma(Z_k) > 3$ are shown. Curves show the variation of $R_k$ with optical depth (top axis)
for $X_{iso}$=30 (dashed lines) and 60 (solid lines); $T_x(^{12}{\rm CO})$=8~K,  $T_x(^{13}{\rm CO})$=8~K (black lines)
and $T_x(^{12}{\rm CO})$=8~K,  $T_x(^{13}{\rm CO})$=6~K (gray lines).
}
\label{figure10}
\end{center}
\end{figure*}

To compile $R_k$ from the data, we constructed the mean \co\ and \coa\ spectrum from all pixels in a subfield
and aligned these spectra to the same spectral axis.  Then, the observed ratio for that subfield is 
$R_k = \left< T_{k}(^{13}{\rm CO}) \right>/\left<T_{k}(^{12}{\rm CO})\right>$.  Random errors are propagated
to derive $\sigma(R_k)$. We consider the same set of velocity channels as those shown in Figure~\ref{figure6}. 
Figure~\ref{figure10} shows the variation of observed $R_k$ values with velocity displacement with points colored by 
their $Z_k$ value. For clarity, we only show points with significant ($>3\sigma$) values of $R_k$ and 
$Z_k$.  
Curves of equation 11 are shown with opacity along the top axis for 2 values 
of $X_{iso}$ = 30 (dashed) and 60 (solid) 
and $T_x(^{13}{\rm CO})$=$T_x(^{12}{\rm CO})$=8~K (black) and $T_x(^{13}{\rm CO})$=6~K, $T_x(^{12}{\rm CO})$=8~K (gray) to show the effects of 
isotopic fractionation and subthermal excitation.
Both plots show similar patterns.  There is a broad distribution of points at low values of $R_k$ and 
a convergence of
$R_k$ towards its peak at zero velocity displacement.
Subfields with positive $Z_k$ values indicating gradient normals mostly aligned with 
the magnetic field orientation have low values of $R_k$ and optical depths that are low but still greater than 1.  
While the positive $Z_k$ subfields are distributed across the 
range of velocity displacements, there are more subfields with positive velocity displacements as noted in 
\S5.2.  
Subfields with negative $Z_k$ values show a much broader distribution of $R_k$ values.  In the right panel of 
Figure~\ref{figure10}, the domain defined 
by $R_k > 0.3$ is almost exclusively populated by subfields with $Z_k(^{13}{\rm CO}) < 0$. 

The line opacity limits the depth to which our observations can probe over a given velocity interval.  The high opacity of \co\
implies that one can detect gas layers at different velocities but not penetrate very deeply into a given layer.  
The surplus of positive $Z_k(^{12}{\rm CO})$ and $Z_k(^{13}{\rm CO})$ values with low $R_k$ suggests that reduced 
optical depths are 
necessary to measure surface brightness textures and recognize magnetically aligned features in molecular clouds. 
In contrast, subfields with thin slice intensity gradient normals mostly perpendicular to the magnetic field orientation (blue points) show a broader distribution of 
$R_k$ values corresponding to a larger range in line opacities for both \co\ and \coa.  
The visibility 
of such features over a range of $R_k$ suggests that such orientations are 
less influenced by line opacity. 

\section{Discussion}
Constraining the role of the interstellar magnetic field in molecular clouds is of paramount importance to our 
understanding of cloud evolution and star formation.  A strong magnetic field with respect to a cloud's self-gravity can 
slow the rate of collapse to account for the observed inefficiency of star formation \citep{Mouschovias:1987}.  
Furthermore, the 
local Alfv\'enic Mach number, M$_{\rm A}$,  
can influence the development of filaments, clumps, and cores and the topology of the magnetic field orientations. 

Our images of $Z_k$ and $Z_{int}$ are roadmaps to the influence of the magnetic field upon the gas.
We identify distinct regions in the Taurus cloud in which 
brightness temperature gradients are aligned parallel or perpendicular to the magnetic field orientations
depending on location and line 
opacity (\co\ or \coa).  We also find subfields where there is no preferred alignment. 

The strong, parallel alignment of brightness temperature gradients with the magnetic field orientation,
as in the northeast sector of the map, indicates that the 
motions are sub-Alfv\'enic \citep{Lazarian:2018}.  In this regime,
the magnetic field configuration is not distorted by the isotropic turbulent motions.
Rather, the gas 
motions are anisotropic and regulated by the magnetic field threading the cloud.   

Our study confirms that the relative orientation becomes mostly perpendicular in subfields with hydrogen 
column densities in excess of 6$\times$10$^{21}$ cm$^{-2}$.  
This is likely due to the effects of gravitational collapse in 
these dense regions \citep{Chen:2016, Soler:2017}.
If the 
dense filaments, which comprise much of the central region of the Taurus cloud, 
develop from 
material falling along a uniformly configured magnetic field due to gravity, then 
this naturally sets up perpendicular 
orientations of the \coa\ brightess temperature gradients and the magnetic field orientation. 
Alternatively, large shocks driven by localized, supersonic yet sub-Alv\'enic flows
channeled by the magnetic field would also generate filamentary structures perpendicular to 
the magnetic field \citep{Chen:2015,Mocz:2018}.   


For both isotopologues, 33\% to 50\% of the subfields have $|Z_{int}| <$ 3 and no preferred orientation.
In these areas, there is either 1) little signal, or 
2) if any signal is present,
the surface brightness in the thin slice channels is uniformly distributed,  so there are no gradients
 or 3) the orientations of 
thin slice intensity gradients 
are randomly offset
from the magnetic field orientations.  Super-Alfv\'enic gas flows can be responsible for 
the last of these conditions as isotropic turbulence can scramble the orientations of the magnetic field.
However at the 10\arcmin\ resolution of the Planck measurements, the magnetic field orientations vary smoothly across most 
of the Taurus cloud 
indicative of sub-Alfv\'enic gas motions.  So the absence of a preferred orientation is produced 
by the broad distribution of 
orientations of the intensity 
gradients with respect to a mostly uniform field and not to a distorted magnetic field. It is possible that the 
gas motions in these subfields are trans-Alfv\'enic, M$_{\rm A}$ $\sim 1$,  
such that the magnetic 
field configuration is not strongly distorted but the gas motions are less responsive to the local magnetization 
\citep{Heyer:2012}.  
This conjecture is supported by \citet{Hu:2019} who derive an Alfv\'enic Mach number in Taurus of 1.1-1.2
by applying the Davis-Chandrasekhar-Fermi method 
to the distributions of Planck polarization angles and \coa\ velocity centroid gradient angles.

Our analysis identifies subfields with intensity gradients parallel and perpendicular to the magnetic field orientation as well as 
subfields with no preferred alignment. These differential conditions  suggests that the Alfv\'en Mach number varies within a cloud volume.  
Such inhomogeneity should be  expected due to strong,
localized, density fluctuations in a supersonic medium and the $\rho^{-1/2}$ dependency, where $\rho$ is the mass volume density.
Given the density inhomogeneity of molecular clouds, 
there are necessarily pockets of large and small Alfv\'enic Mach numbers 
\citep{Burkhart:2009}. 
This is an important observational insight, as 
it suggests to simulators and 
theorists that there may not be one magnetic state within a cloud domain. Consequently, 
there should be a range of possible star formation 
scenarios within the domain of a single cloud. 

\section{Conclusions}
We have examined the relative orientations between the interstellar magnetic field and spatial derivatives of antenna 
temperatures within thin velocity slice images of \co\ and \coa\ J=1-0 emission from the Taurus molecular cloud. 
Our analysis includes a robust treatment of random errors in the relative orientations produced by the thermal noise of the 
spectroscopic and polarization data.  The relative orientations within subfields of the cloud are parameterized 
by the projected Rayleigh statistic and the mean resultant vector that are derived for both thin-sliced velocity 
channels and integrated over a broader velocity interval.  
Images of the projected Rayleigh statistic show 
mostly parallel alignments of the vector normal to the intensity gradient and the component of the
interstellar magnetic field in the plane 
of the sky for subfields in the outer periphery of the cloud and in subfields with reduced \co\ optical depths. 
This relative orientation becomes mostly perpendicular in 
subfields in the central regions of the cloud with hydrogen column densities greater than 6$\times$10$^{21}$ cm$^{-2}$.  The 
strongest alignments (parallel or 
perpendicular) occur in regions for which measures of the amplitude of turbulent motions are comparable to values 
found in Galactic clouds.   There are other sections of the cloud where there is no preferred alignment of gradients 
with the magnetic field that may imply a transition to trans-Alfv\'enic Mach numbers but could also be a result of 
projection effects 
along the line of sight or inefficient  grain alignment with the local magnetic field.
This diversity of alignments imply changing levels of magnetization within the cloud volume.  Such inhomogeneity 
should be considered when addressing the evolution of molecular clouds and the formation of stars. 

\section*{Acknowledgments}
The authors acknowledge the anonymous referee whose suggestions improved the clarity of this manuscript.
J.D.S is funded by the European Research Council under the Horizon 2020 Framework Program via the ERC Consolidator Grant CSF-648505.
All three authors acknowledge Paris-Saclay University's Institut Pascal program "The Self-Organized Star Formation Process" and the 
Interstellar Institute for hosting discussions that nourished the development of the ideas behind this work.
This research made use of Astropy,\footnote{http://www.astropy.org} a community-developed core Python package for 
Astronomy \citep{astropy:2013, astropy:2018}.

\section*{Data Availability}
The data underlying this article will be shared on reasonable request to the corresponding author.

\bibliographystyle{mnras}
\bibliography{cite_MHDTaurus.bib} 

\appendix
\section{Instrumental Errors}
The analysis of spatial gradients of two-dimensional fields is a powerful tool that quantifies the rich structure and 
texture of molecular line emission or thermal dust emission. However, the derived gradients of observational data are sensitive to 
the random errors of the measurements.  For line emission, uncertainties of observed brightness temperatures arise from 
fluctuations of the radiometer and those of the atmosphere through which the measurements are collected.  The root 
mean square (rms) fluctuations due to thermal noise can 
be derived for each spectrum by simply calculating the standard deviation of values in spectral channels with no emission.
These errors propagate through the calculation of the gradient magnitude and direction (equations 1 and 2).  To assess the 
impact of instrumental errors on the gradient vector, we use the Monte Carlo method for its simplicity.  

\begin{figure*}
\begin{center}
\epsfig{file=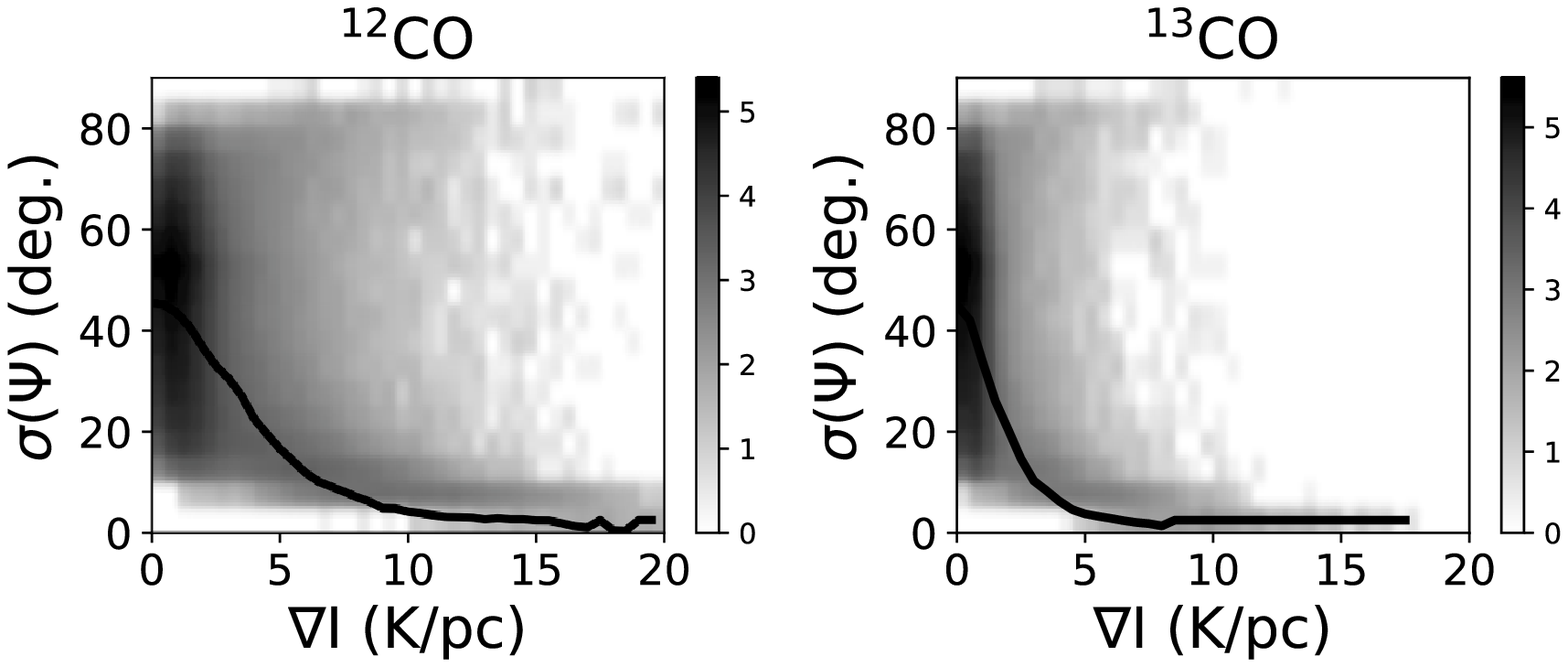,width=17cm,clip=}
\caption{A two-dimensional histogram of random errors of the gradient angle, $\sigma(\Psi)$ and the magnitude of the intensity gradient, $\nabla$I, for 
\co (left) and \coa\ (right).  The shading of the histogram is log(N), where N is the number of points in each two-dimensional
bin.  
The solid line shows the median value of $\sigma(\Psi)$ for each bin of $\nabla$I.  The gradient angle, $\Psi$ is 
only well defined ($\sigma(\Psi) < 15^\circ$) for gradients larger than 4 K/pc (\co) and 2.2 K/pc (\coa). 
}
\label{figureA1}
\end{center}
\end{figure*}

We summarize the Monte Carlo derived errors in $\Psi_{G,k}$ in Figure~\ref{figureA1}, which shows a two dimensional histogram in bins
of $\sigma(\Psi_{G,k})$ and the gradient magnitude, $\nabla T_k$ for both \co\ and \coa.  Each histogram is constructed from almost 16 
million values corresponding to 1024 samples for each subfield, 150 subfields, and 104 spectral channels.  
For each $\nabla T_k$ bin, 
we derive the median value of $\sigma(\Psi_{G,k})$, which is shown as the solid, dark line in Figure~\ref{figureA1}.  
The median values decrease with 
increasing gradient magnitude.  
For small gradient 
magnitudes, the errors in the gradient direction are quite large as expected 
for a random distribution.
For display purposes only (see Figures~2,3), we determine the value of the gradient magnitude at which $\sigma(\Psi_{G,k})$ is less 
than 15$^\circ$ for each CO isotopologue. These threshold values are 4 K/pc and 2.2 K/pc for \co\ and \coa\ respectively.

\begin{figure}
\begin{center}
\epsfig{file=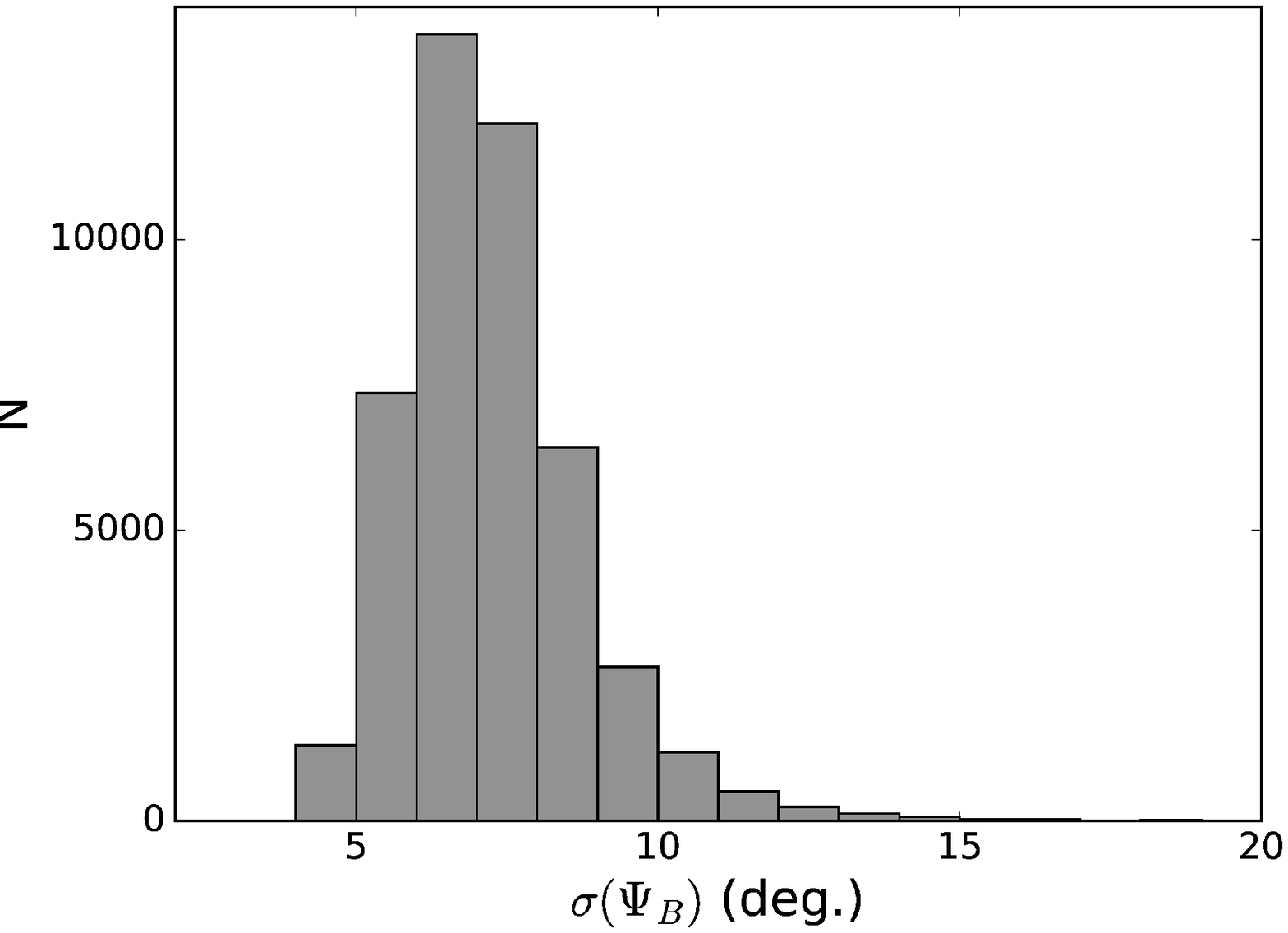,width=9cm,clip=}
\caption{A histogram of the instrumental errors of the magnetic field orientation, $\sigma(\Psi_B)$ as derived from the uncertainties of
the Planck 353~GHz $U$ and $Q$ Stokes parameters. The median error is 7$^\circ$.  
Random errors of the relative orientation, $\sigma(\Phi)$ are 
dominated by those of the gradient. 
}
\label{figureA2}
\end{center}
\end{figure}

Instrumental uncertainties also affect the polarization angle, $\Psi_{E}$ from which we infer the orientation 
of the interstellar magnetic field assuming the alignment of elongated grains. The error in the polarization 
angle is derived from the errors in the $U$ and $Q$ Stokes parameters.  We do not add any error terms in the 
rotation of the polarization angle by 90$^\circ$ to derive $\Psi_B$.
The distribution of errors in $\Psi_B$ are shown 
in Figure~\ref{figureA2}.  The mean error is 7.3$^\circ$ with a standard deviation of 1.6$^\circ$.  

The resultant errors in 
the gradient direction as well as errors in the dust polarization orientation that arise from uncertainties in the $U$ and $Q$
Stokes parameters are added in quadrature to determine the error in the relative orientation, $\sigma(\Phi)$ for each sampled 
position.  
The errors in the 
gradient calculation make the largest contribution to the uncertainties of the relative orientation, $\Phi_k$, 
for all but the largest gradients.
This composite error value is used when weighting the sums for the Rayleigh statistic and the mean resultant vector. 
For the thin-slice channels with no emission, the $Z_k$ values are all near zero as expected for a 
random distribution of angles derived from noise as shown for a single subfield in Figure~\ref{figure5}. 
Moreover, the random errors in $Z_k$ derived from propagating the Monte-Carlo produced distributions of $\Psi_{G,k}$,
 are comparable to 
the predicted values \citep{Jow:2018}. 

\section{Correlation between $r_{int}$ and $Z_{int}$}
\begin{figure}
\begin{center}
\epsfig{file=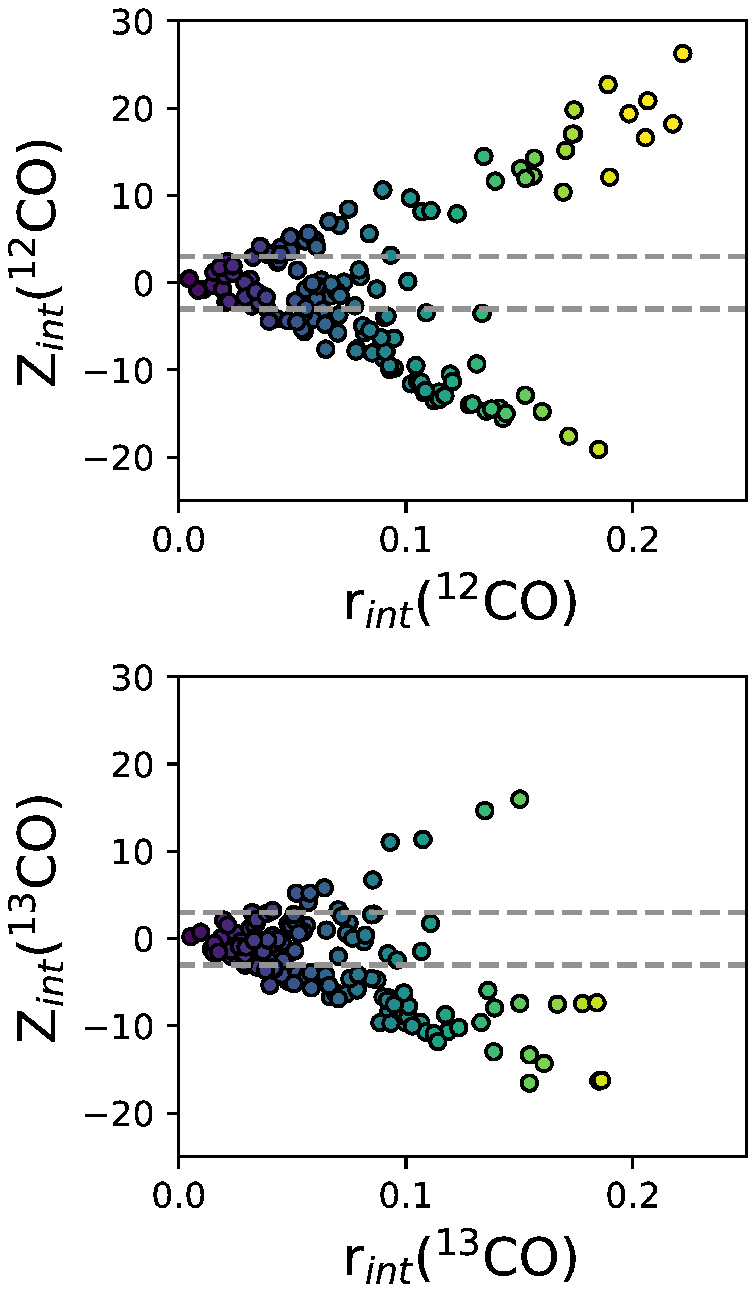,width=6cm,clip=}
\caption{Variation of the Rayleigh statistic, $Z_{int}$, with the mean resultant vector, $r_{int}$ for \co (top)
and \coa (bottom).  The dashed horizontal lines denote 3$\sigma(Z_{int})$ values. 
The color coding of points is the same as in Figure~\ref{figure7} and 
Figure~\ref{figure8}.  $Z_{int}$ correlates well with $r_{int}$ for $r_{int} > 0.1$.
}
\label{figureB1}
\end{center}
\end{figure}

In \S4.2, we use the mean resultant vector, $r_{int}$ as a guide to 
interpret the signficance of $Z_{int}$ values. 
In Figure~\ref{figureB1}, we show the variation of $Z_{int}$ values 
with $r_{int}$ for all 150 subfields 
to evaluate the connection between the two statistics and 
to define a threshold value of $r_{int}$ above 
which one can have confidence in the $Z_{int}$ value.  The dashed horizontal lines denote $|Z_{int}|/\sigma(Z_{int}) = 3$
Within the 3$\sigma$ limits, there is a cluster of points with $r_{int} < 0.05$ that arises from subfields with little 
or no signal.  For $r_{int}> 0.1$, the two statistics correlate very well.  Examining 
Figure~\ref{figure7}, one should place more confidence in $Z_{int}$ values with $r_{int}$ values above this threshold. 

\section{Resolution Study}
For our study, the angular resolution of the Rayleigh statistic is set by the size of the subfields.  
In Figure~\ref{figure6} and Figure~\ref{figure7}, each 
subfield is comprised of 256$\times$256 pixels and subfields are spaced from each other by 128 pixels.  
When calculating the Rayleigh statistic (equations 4 and 8) and the mean 
resultant vector (equations 9-11), the subfield is sampled at 8 pixel intervals, which provides 
1024 relative orientation angles for each subfield.  However, as Figure~\ref{figure2} demonstrates, there are 
temperature brightness variations in the \co\ and \coa\ thin-slice images.  Small scale 
variations of $\Phi$ values could impact the value of the Rayleigh statistic 
when summed over a given area. 

\begin{figure*}
\begin{center}
\epsfig{file=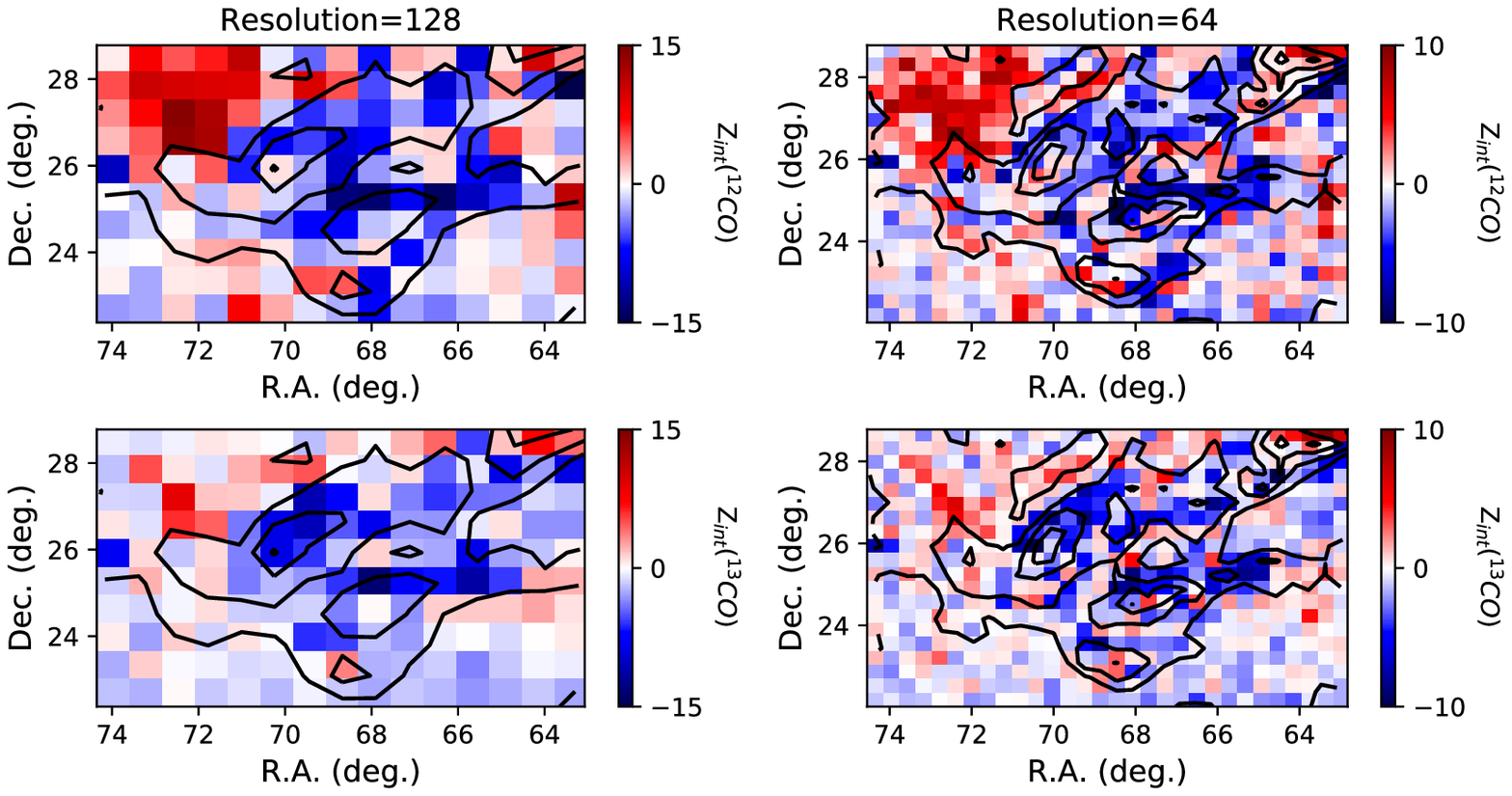,width=17cm,clip=}
\caption{Images of the Rayleigh statistic, $Z_{int}$, for resolutions of 128 pixels (left) and 
64 pixels (right) for both \co\ (top) and \coa\ (bottom). 
Black contours show the distribution of hydrogen column density at these same resolutions.
}
\label{figureC1}
\end{center}
\end{figure*}

To assess the effects of resolution on the value of $Z_{int}$, we have constructed images of the 
Rayleigh statistic within subfields sizes of 128$\times$128 pixels and 64$\times$64 pixels.
The subfields are spaced by 128 and 64 pixels respectively.  These resolutions provide 
256 and 64 samples for each subfield.   The results are shown 
in Figure~\ref{figureC1} for both \co\ and \coa.   
In general, the $Z_{int}$ images at these higher resolutions are similar to those shown in Figure~\ref{figure7}
but with lower values of $Z_{int}$. 
Since $Z_{int}$ is an additive, unnormalized quantity, 
the absolute value of $Z_{int}$ decreases with decreasing number of $\Phi_k$ values.  

\begin{figure*}
\begin{center}
\epsfig{file=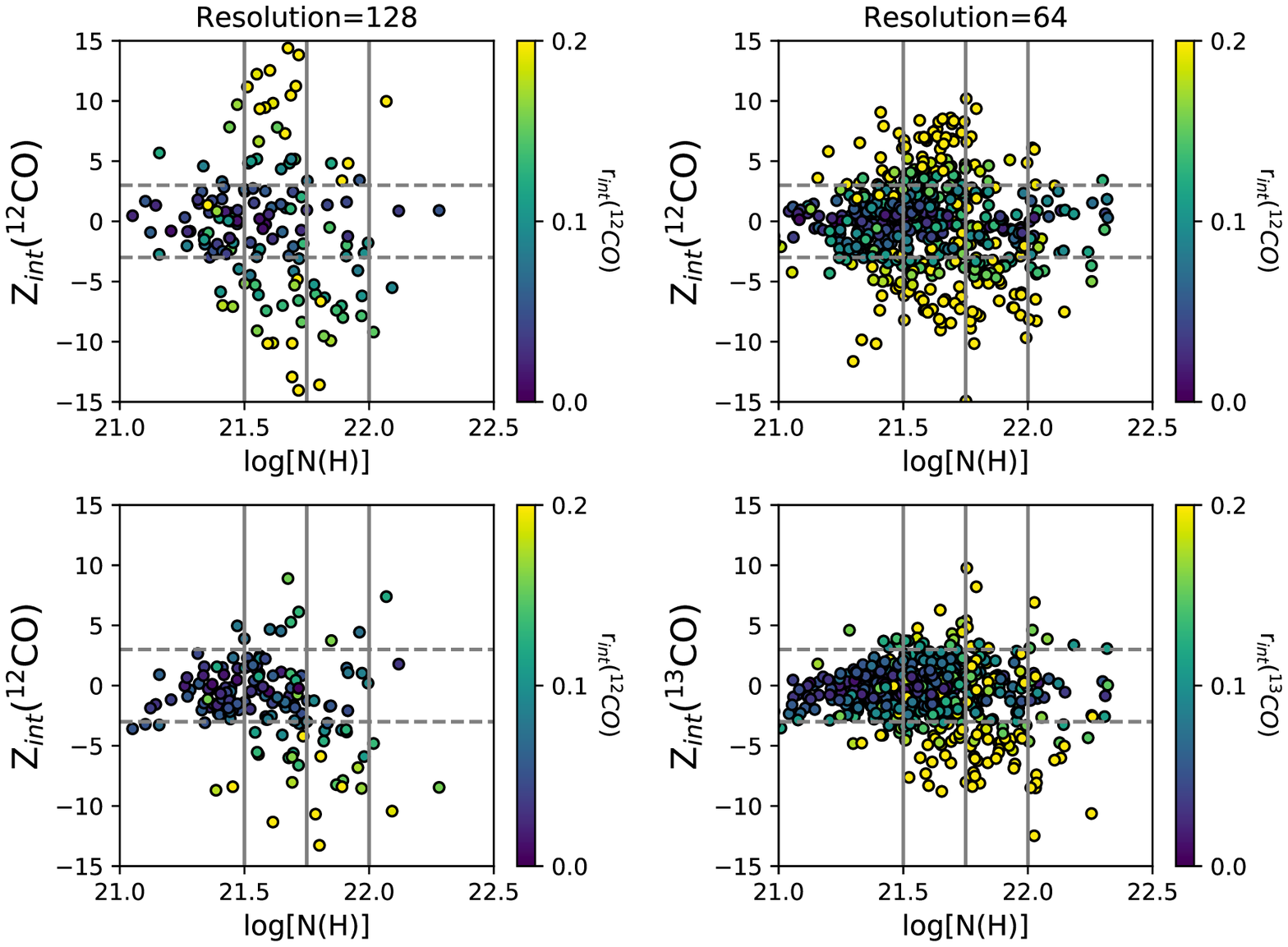,width=15cm,clip=}
\caption{Variation of $Z_{int}$($^{12}${\rm CO}) (top) and $Z_{int}$($^{13}${\rm CO}) (bottom) with hydrogen column density 
in each subfield with
resolutions of 128 pixels (left) and 64 pixels (right). 
The color coding of points reflect $r_{int}$ values calculated with these same resolutions. 
}
\label{figureC2}
\end{center}
\end{figure*}

Since the Taurus cloud exhibits strong spatial variations in hydrogen column density, the lower resolution 
information on $Z_{int}$ and N$_H$ could smooth over small scale correlations between these 2 quantities. 
Figure~\ref{figureC2} shows the variation of $Z_{int}$ values with hydrogen column density for these 
higher resolutions.  Similar trends that are identified in 
Figure~\ref{figure8} and Figure~\ref{figure9} are also evident at these higher resolutions but extend to higher column 
density values owing to the finer resolution.  No new trends emerge from the higher resolution images. 

\end{document}